\documentclass[reprint,aps,prl,twocolumn,superscriptaddress]{revtex4-1}
\usepackage{graphicx}
\usepackage{mathrsfs}
\usepackage{bm}
\usepackage{hyperref}
\usepackage{dcolumn}
\usepackage{amsmath}
\usepackage{amssymb}
\usepackage{color}
\usepackage{bbold}
\usepackage{braket}
\hypersetup{  colorlinks=true, linkcolor=black, citecolor=black, urlcolor=black  }
\usepackage{ifpdf}

\newcommand{\br}{{\mathbf r}}
\newcommand{\bk}{{\mathbf k}}

\newcommand{\bK}{{\mathbf K}}
\newcommand{\bR}{{\mathbf R}}
\newcommand{\bG}{{\mathbf G}}

\newcommand{\bL}{{\mathbf L}}

\begin{document}
\title{Silicon Donor Array as a Disordered One-Dimensional Electron Gas}
\author{Chao Lei}
\affiliation{Department of Physics, The University of Texas at Austin, Austin, Texas 78712,USA}
\author{Allan H. MacDonald}
\affiliation{Department of Physics, The University of Texas at Austin, Austin, Texas 78712,USA}

\begin{abstract}
Donors in silicon can now be positioned with an accuracy of about one lattice constant, making it
possible to form donor arrays for quantum computation or quantum
simulation applications. However the multi-valley character of the silicon conduction band combines
with central cell corrections to the donor state translate atomic scale imperfections in donor placement into strongly disordered 
inter-donor hybridization.  We present a simple model that is able to
account for central-cell corrections accurately, and use it to assess the impact of donor positional
disorder on donor array properties in both itinerant and localized limits.  
\end{abstract}

\maketitle

\textit{Introduction---}
One strategy for establishing robust solid-state quantum information processing hardware is to 
exploit the relatively simple bound states
that surround donors or acceptors in the best understood semiconductor material, silicon \cite{Rev_silicon2013,Rev_DonorQC2003}.
Considerable experimental progress has been made toward the use the electron spin of a donor-bound state in silicon,
or alternately the donor nuclear spin, as a qubit \cite{qubit2012,readout2013,Lo2013,Pla2014,Kalra2014,Sigillito2017,Gates2015,He2019,Morsee2017,Hile2018,Dehollain2014,Broome2017,Zajac2018,Watson2018,Broome2018,Wang2016}.
It has been possible, for example, to achieve long coherence times for both electron \cite{coherence_e2011,Watson2017} and nuclear spins\cite{coherence30s2014}.
Additionally, advances in the technology for deterministically implanting donors in silicon \cite{Mott2012,2D_gas2013,donor2d2014,donorchain2016,Cooil2017}
with high positional accuracy have made it possible to form donor arrays, 
which are attractive for both quantum computation \cite{Kane1998} and quantum simulation \cite{Mott2012,Salfi2016,Ansaloni_2020} applications.

The physics of donor arrays in silicon is complicated by the presence of six valleys in the silicon conduction band,
which adds an unwanted valley degree-of-freedom to donor-bound electron envelope function Hamiltonians.  
The valley degeneracy is
fortunately lifted by central-cell corrections that couple valleys \cite{Valley_in_silicon2012,Goh2020}.
There is still an unwanted complication, however, since mixing between valleys makes the interactions between
donor levels extremely sensitive to donor positional disorder, as we will discuss in this paper.  Even 
the lattice-constant-scale accuracy in donor positioning, now achievable
in a silicon crystal \cite{Precise2003}, is not necessarily adequate.
To describe this physics,
we introduce a model for the central cell interactions that are responsible for valley splitting of donor levels in bulk silicon.  The model is attractively simple and captures all the essential multi-valley physics.
We use it to assess the effectiveness of recently proposed strategies \cite{Salfi2018,Voisin2020} to 
mitigate the influence of positional disorder on the exchange coupling between two spin qubits
in the localized limit, and on donor array band properties in the itinerant limit.

Below we first present a calculation of valley-dependent tunneling 
in one-dimensional donor arrays, and then combine it with our simplified theory of valley-mixing by 
central cell corrections to quantify donor array disorder.  In the localized limit,
only donor spin degrees of freedom are relevant.   We therefore use our model to calculate 
the exchange interactions between neighboring spins, demonstrating that they are
less sensitive to positional disorder when oriented along $\langle 110 \rangle$ 
rather than along $\langle 100 \rangle$, as demonstrated recently ~\onlinecite{Salfi2018,Voisin2020}.
In the itinerant limit, positional disorder localizes donor array Bloch states.  We show that 
this effect is also strongly limited by placing the donor array along $\langle 110 \rangle$.
Our calculations demonstrate that central-cell corrections which separate the 
$A_1$ bound state from other disorder levels play a central role in determining donor-array
properties in both localized and itinerant limits.

\textit{Donor array model}---
Donors in silicon have been well understood for decades \cite{Theory_Kohn1955,Theory_Luttinger1955,Theory_Kittel1954,Shallow1973,Shindo1976,Saraiva2011}.
Here we employ an effective-mass approach \cite{Theory_Kohn1955,Theory_Luttinger1955} in which the wavefunction of an electron
bound to an isolated donor has the form:
\begin{equation}
  \psi(\br) = \sum_{\mu = 1}^{N_{\mu}} F_{\mu}(\br) \phi_{\mu} (\br)
\end{equation}
where $\mu$ labels valley, $N_{\mu}=6$ is the number of valleys, 
$F_{\mu}(\br)$ is an envelope function
and $\phi_{\mu}(\br) = e^{i\bk \cdot \br}u_{\mu}(\br) $
is a band minimum Bloch function with periodic factor $ u_{\mu}(\br) $.
When central cell corrections are neglected there is no coupling between valleys and
donor-bound states in each valley are eigenstates of an effective mass Schr\"{o}dinger
equation\cite{Theory_Luttinger1955} with Hamiltonian:
\begin{equation}\label{shodinger}
 H = \sum_i {\frac{\hbar^2}{2m_i^{\ast}} \nabla_i^2 + V(\br) }.
\end{equation}
Here $i = x,y,z$, $m_i$ are effective masses, and $V(\br) =  -e^2/\epsilon |\br|^2$ is the hydrogenic external potential
induced by the replacement of a Si atom by a donor ion at the origin.  The mass tensor in Eq.~\ref{shodinger}
is diagonal because the six conduction band valleys in silicon are located
along the principle cubic axes, with a large mass $m_i$ for momentum along the valley direction
(longitudinal mass $m_l \approx 0.98 \, m_0$ where $m_0$ is the bare electron mass ) and
a small mass for perpendicular momentum (transverse mass $m_t \approx 0.19 \, m_0$).

Because we are interested in a periodic array of donors, we use a plane-wave expansion approach 
and place donors at the center of three-dimensional supercells with dimension $L_x \times L_y \times L_z$.
The donor array envelope functions then depend on
wavevector $\bk$ in the array mini-zone and can be expanded in the form:
\begin{equation}\label{fmu}
 \ket{F_{\mu}(\bk)} = \sum_{\bG}{C_{\bG}(\bk) \ket{\bk+\bG} },
\end{equation}
where $\bk$ is the wavevector and $\bG$ is the supercell reciprocal lattice vectors (see \cite{supplement} for more details).
The kinetic energy is dependent on the spatial orientation of the donor array relative to the cubic axes.
We limit our attention to donor arrays that 
have their $\hat{z}$-axis aligned with the silicon $\hat{z}$ direction,
but allow for changes of orientation $\theta$  of the donor array $\hat{x}$ and 
$\hat{y}$ directions relative to the silicon crystal $\hat{x}$ and 
$\hat{y}$ axes. 
For this family of donor array orientations 
\begin{equation}
T_{\mu\mu^{\prime}}^{\bG\bG^{\prime}} = \delta^{\bG\bG^{\prime}}_{\mu\mu^{\prime}} \sum_{i} \frac{\hbar^2}{2m_{i}^{\ast}} \Big(\sum_j (k_j + G_j) R_{ij}(\theta)\Big)^2 ,
\end{equation}
where $ \delta^{\bG\bG^{\prime}}_{\mu\mu^{\prime}} \equiv \delta_{\mu\mu^{\prime}}\, \delta_{\bG\bG^{\prime}} $ and $R_{ij}(\theta)$ is the rotation matrix.
The external potential of the donor array\cite{Giuliani2005} is that produced by 
unit positive charges at the center of each supercell so that
\begin{equation}
V^{\bG\bG^{\prime}}_{\mu,\mu'}(\bk) = -\frac{1}{\Omega} \delta_{\mu\mu^{\prime}} \frac{4\pi e^2}{\epsilon |\bG - \bG^{\prime}|^2},
\end{equation}
where $\Omega$ is the supercell volume and 
$\epsilon \approx 11.7$ is the static dielectric constant of silicon.

As we show explicitly in Fig.~S1 of the supplemental material \cite{supplement}, when all dimensions of the
supercell approach infinity, the envelope functions approach hydrogenic wavefunctions
and the donor binding energies approach $Ry = R_H m^{\ast}/\epsilon^2$ 
where $R_H \approx 13.6 eV$ is the electron binding energy of hydrogen in vacuum, 
and $m^{\ast} \equiv (m_t^2m_l)^{1/3}$ is the conduction band effective mass.
In silicon $Ry \approx 32.6 meV $ and the effective Bohr radius $a_{_{B}} = a_0 \epsilon/m^{\ast} \approx 1.9 nm$.
In the one-dimensional case, the donor states form one-dimensional bands, the band dispersion is strongly valley-dependent as illustrated in Fig.~\ref{Fig:One}, and sensitive to
the spatial orientation of the donor array relative to the cubic axes for a given valley. 
Since the rotation of donor array happens in $\hat{x}-\hat{y}$ plane, hopping within $\pm z$ valleys 
mass is insensitive to the rotation angle $\theta$, whereas for the $\pm x$ or $\pm y$ valleys, 
hopping on the array depends strongly on the the orientation angle.
For a $\langle 110 \rangle$ orientation $x$-valley and $y$-valley hopping have identical amplitudes that are 
one order of magnitude smaller than for $z$-valley hopping.

\ifpdf
\begin{figure}
  \centering
  \includegraphics[width=0.9 \linewidth ]{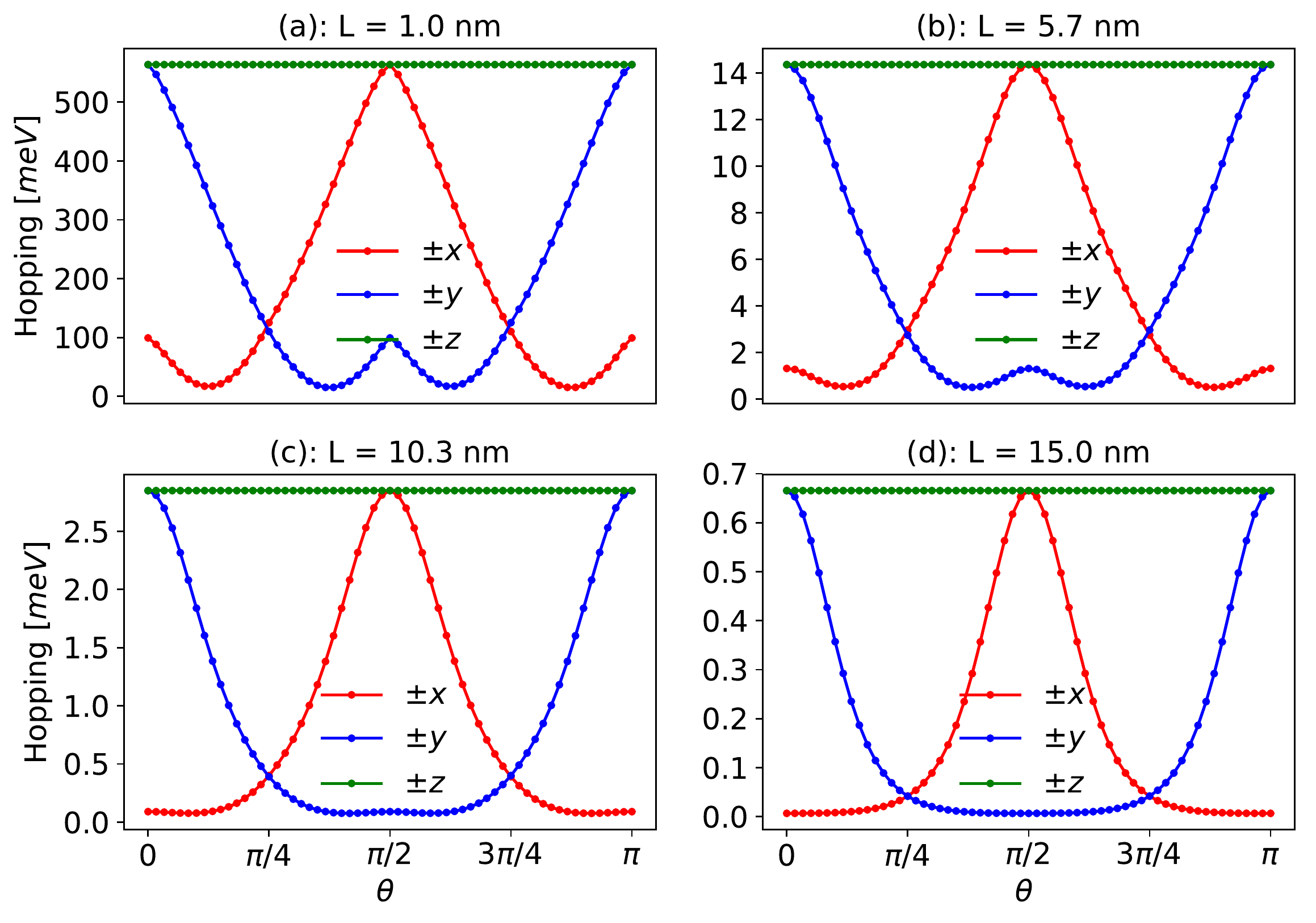}
  \caption{Valley and orientation dependent near-neighbor hopping parameters for 
  cubic donor-arrays with lattice constants $L$.  The $\pm x$, $\pm y$ and $\pm z$ labels 
  specify the Si conduction band valleys centered at 
  $ \bK_{\mu} = \pm k_0 \hat x, \pm k_0 \hat y, \pm k_0 \hat z $, where 
  $k_0 = 0.85 \pi/a$ and $a$ is the lattice constant of silicon.
    }\label{Fig:One}
\end{figure}
\fi

\textit{Central cell corrections}--- 
So far we have neglected the central cell corrections that are important in silicon and yield donor-bound states that couple different valleys.
The binding energies observed experimentally are 45.59 meV for the singly degenerate $1s (A_1)$ level \cite{bindenergy1981,bindenergy1993,Valley_in_silicon2012},
32.58 $meV$ for the doubly degenerate $1s (E)$ level,
and 33.89 meV for the triply degenerate $1s (T_2)$ level.
If we ignore the small difference between the $1s (E)$ and $1s (T_2)$ levels,
we can approximate the central cell correction correction to the single-particle Hamiltonian as the product of
a single valley splitting energy scale $\epsilon_{vs} \approx 12$ meV ~\cite{Valley_in_silicon2012,Multivalley2015}, an attractive delta-function $-\epsilon_{vs} \delta(\br - \bR)$ at the donor site $\bR$,
and a projection onto the donor states:
\begin{equation}
\label{eq:Hvs}
{\cal H}_{vs}(\bR) =  \frac{-\epsilon_{vs}}{N_{\mu}\Omega} \; \sum_{\substack{\bk'\bk \\ \mu'\mu}} \vert \bk'\mu'\rangle \langle \bk\mu \vert \; e^{i(\bK_{\mu}+\bk-\bK_{\mu'} - \bk') \cdot \bR},
\end{equation}
where $ \bK_{\mu} = \pm k_0 \hat x, \pm k_0 \hat y, \pm k_0 \hat z $ are valley-momenta in bulk silicon,
$k_0 = 0.85 (2\pi/a)$, $a = 0.543$ nm is the conventional cubic lattice constant of bulk silicon, 
Because $\bK_{\mu}-\bK_{\mu'}$ is comparable in size to a 
reciprocal lattice vector, the valley splitting Hamiltonian changes substantially even 
for changes in $\bR$ that are on the atomic length scale.  
Note that Eq.~\ref{eq:Hvs} contains $\mu' \ne \mu$ terms that couple different valleys.

In order to calculate the parameters of a generalized Hubbard model for the donor array,
we first Fourier transform the wavefunctions in the lowest energy band back to real space:  
\begin{equation}
 | \bR{\mu \sigma} \rangle = \sum_{\bk, \bG}{C_{\bk\bG}^{\mu \sigma} e^{-i\bk \cdot \bR} | \bk + \bK_{\mu} + \bG} \rangle | \sigma \rangle,
\end{equation}
where $\mu $ labels valley, $\sigma$ labels spin, $\bR$ is the position of the donor,
and we have chosen $C_{\bk\bG=0}$ to be real and positive at each value of $\bk$.
The single-particle Hamiltonian in the Wannier representation defines the 
donor array hopping parameters $t_{\bR\bR^{\prime}}^{\mu\mu'}$,
which are a sum of kinetic enegy ($T_{\bR\bR^{\prime}}$) and external potential ($V_{\bR\bR^{\prime}}^{ext}$) 
contributions (see \cite{supplement} for more details).

In Fig. \ref{Hopping_U} (a) and (b) we plot hopping 
parameters for the $\pm x$ and $\pm y$ valleys 
as a function of donor separation and the spatial orientation of the donor array relative to 
the $x$ axis.  We see here that at small donor distances minima of the hopping 
parameters appear at $\theta \approx \pi/8$ for $\pm x$ valleys and at $\theta \approx 3\pi/8$ for $\pm y$ valleys. While for the $\pm z$ valley, the hopping parameters are independent of $\theta$ and the decay lengths of the hopping parameters extracted from the valley wavefunctions are nearly idependent of $\theta$. 
In the $\langle 100 \rangle$ direction both $\pm y$ and $\pm z$ effective masses are 
transverse ($m_t$) and therefore have the same decay length, which is larger than the 
$\pm x$ valley decay length which has the longitudinal effective mass $m_l$. 
Similar results apply for $\pm x$ and $\pm z$ valley in the $010$ direction. Along 
the $\langle 110 \rangle$ direction ($\theta = \pi/4$) the decay lengths of $\pm x$ and $\pm y$ valleys are identical since 
the wavefunctions have equal contributions from transverse and longitudinal mass components.

\ifpdf
\begin{figure}
  \centering
  \includegraphics[width=0.9 \linewidth ]{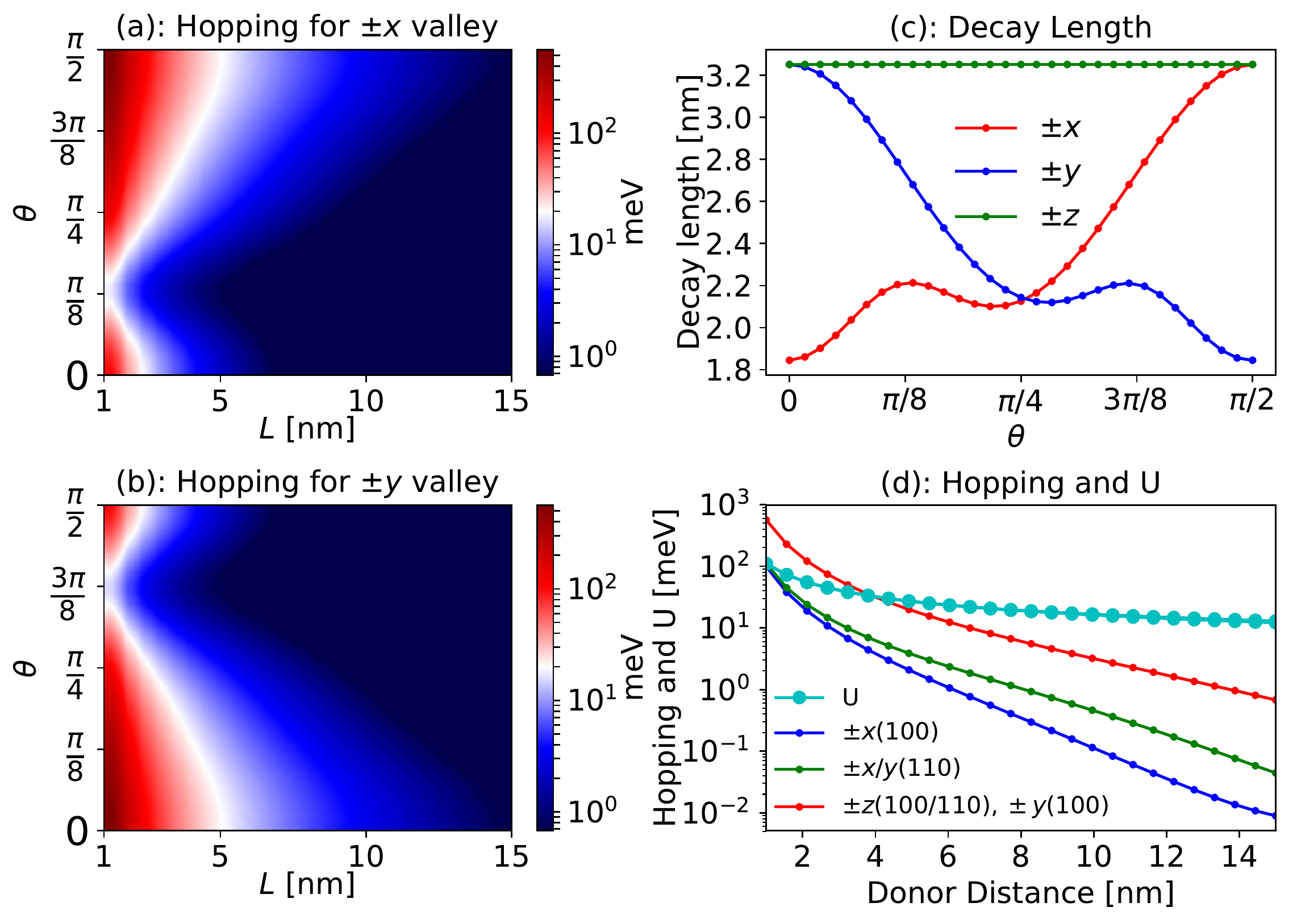}
  \caption{Hopping parameters and on-site interactions of the silicon donor array Hubbard model 
  {\it vs.} the donor separation and the the spatial orientation of the donor array relative to 
  the crystal $x$ axis. The color scale plots in 
  (a) and (b) show the hopping parameter {\it vs.} donor separation $L$ and spatial orientation $\theta$. 
   (c) Decay lengths {\it vs.} $\theta$ for valley $\pm x$, $\pm y$ and $\pm z$.
  (d) Hubbard U and hopping parameters for donor arrays along the $\langle 100 \rangle$ and $\langle 110 \rangle$ directions. 
  }\label{Hopping_U}
\end{figure}
\fi

The on-site Hubbard U can be calculated using $U_{\mu\mu^{\prime}} = \bra{\bR\mu\sigma,\bR\mu^{\prime}\sigma} V_c \ket{\bR\mu\sigma,\bR\mu^{\prime}\sigma}$ where $\ket{\bR\mu\sigma}$ is a Wannier function and $V_c = e^2/\epsilon r^2$.
In Fig. \ref{Hopping_U}(d) we compare hopping parameters for donor arrays along the $\langle 100 \rangle$ and $\langle 110 \rangle$ directions
At small donor separation, valleys with transverse mass have larger hopping parameters 
that are large compared with U.  However on-site electron-electron interaction strengths exceed 
the hopping parameters at larger donor separations; the ratio reaches $\sim 10$ when the donor 
separation is around $12$ nm.

\textit{Valley Interference in Exchange Interactions---}
Using the Hubbard model parameters discussed above, we now assess the influence 
of valley degeneracy and donor placement on the exchange interactions between neighboring sites 
of a donor array.  
To calculate the exchange interactions we study a two-site Hubbard model which includes hopping, valley splitting, and on-site electron-electron interactions
terms:
\begin{equation}
\begin{split}
  H =  & \sum_{\nu} t_{\mu} (c_{1\mu\sigma}^{\dagger}c_{2\mu\sigma} + h.c.) 
  +  \epsilon_{vs} \sum_{i,\mu\mu'} P_{\mu\mu'}^{i}  c_{i\mu\sigma}^{\dagger} c_{i\mu'\sigma}\\
  &+   \sum_{i\mu\mu'\sigma\sigma'} U_{\mu\mu'}  c_{i\mu\sigma}^{\dagger} c_{i\mu\sigma} c_{i\mu'\sigma'}^{\dagger} c_{i\mu'\sigma'}.
\end{split}
\end{equation}
Here $t_{\mu}$ is the inter-site hopping amplitude
within valley $\bK_{\mu}$, $c_{i\mu\sigma}^{\dagger}$ ($c_{i\mu\sigma}$) is a creation (annihilation) operator,
$P_{\mu\mu^{\prime}}^i = e^{i(\bK_{\mu}-\bK_{\mu'}) \cdot \bR}$ is the operator that applies a central-cell energy shift at site $\bR$ to the donor state, and $U_{\mu\mu^{\prime}}$ is the on-site electron-electron interaction, which can be accurately modelled as valley independent. 
For one electron per donor, the charge excitation sector is gapped and low energy states are formed from spin degrees on each site. 
The exchange interaction between spins can therefore be defined in terms of the energy difference between the lowest energy two-electron singlet and triplet states: $J = E_T - E_S$.

\ifpdf
\begin{figure}
  \centering 
  \includegraphics[width=0.9 \linewidth ]{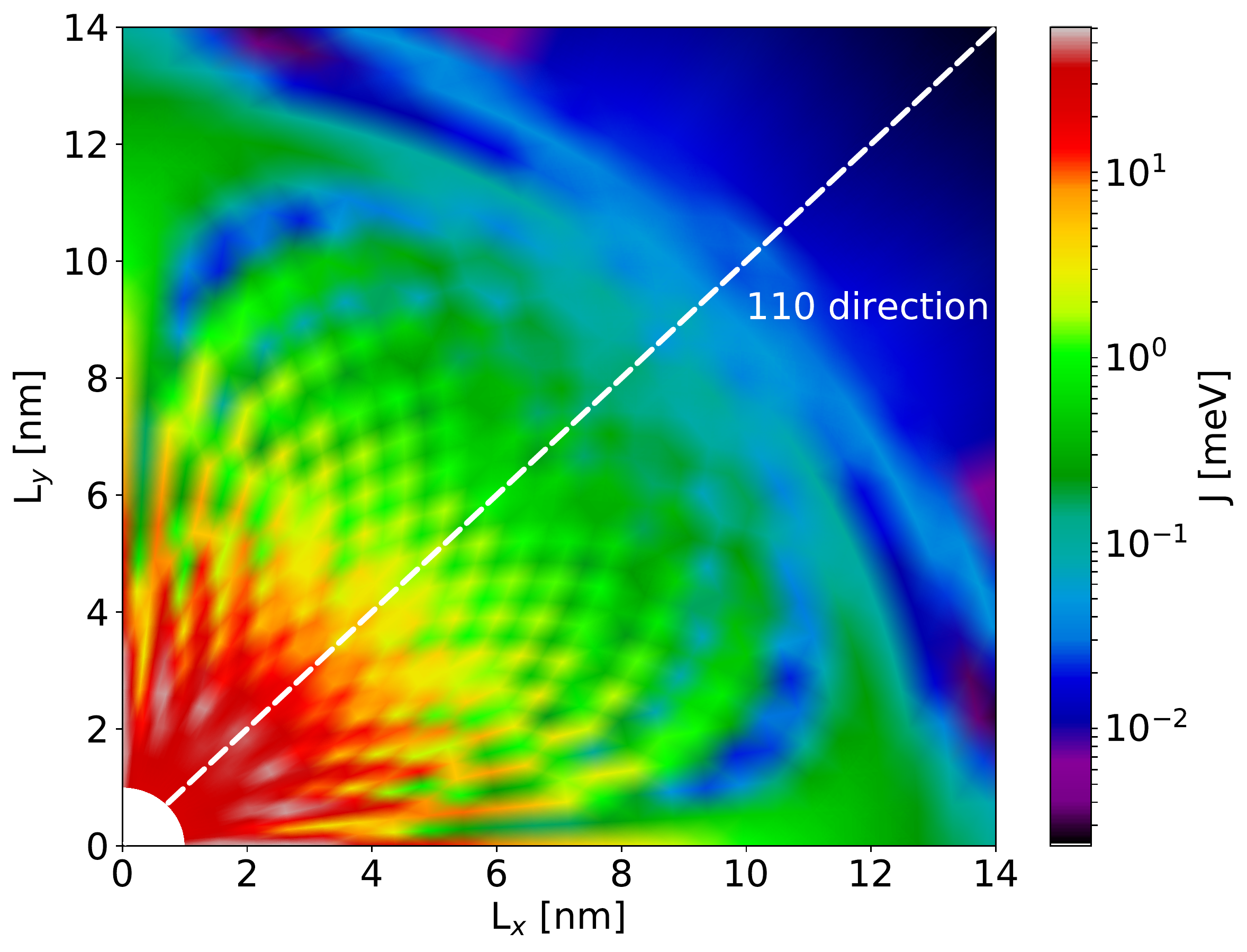}
  \caption{Exchange interaction between two donors calculated with a two-site six-orbital Hubbard model. 
  In explicit calculations, we place one of the two donors at the origin and the other at $\bL = L (1, 0,0)$, where $L$ is the 
  the distance between two donors, and rotate the host-crystal cubic lattice by angle $\theta$. The mapping from a 
  $L-\theta$ grid to the plotted $L_x-L_y$ grid leads to some (white) regions in which there is no data. 
  The white line labels the $\langle 110 \rangle$ direction.
  }\label{Exchange_Interaction}
\end{figure}
\fi

In Fig. \ref{Exchange_Interaction} we plot the exchange 
interaction {\it vs.} donor separation magnitude and direction.  
In our explicit calculations we place one of the two donors at the origin and the 
other at $\bL = L (1, 0,0)$, where $L$ the distance between two donors,
and rotate the host-crystal cubic lattice by angle $\theta$.
The $\pm z$ valleys always have a transverse effective mass $m_t$,
which leads to large inter-site hopping amplitudes.
Because the displacement $\bL$ is perpendicular to the $\pm z$ valley momenta $\bK_\mu$,
the phase factors in the $(\pm z, \pm z)$ blocks of the 
valley-splitting Hamiltonian always vanish ( i.e. $\hat{z} \cdot \bL = 0$, with $\bL \equiv \bR - \bR^{\prime} $).
For general $\theta$ both the $\pm x$ and $\pm y$ blocks of the valley splitting Hamiltonian 
at $\bL$ have non-trivial phase factors, which change in value when $L$ changes on an atomic length scale,
and appear in the hopping amplitude between the two $A_1$ exciton levels.
For $\theta=0$ ($\theta=\pi/2$), $\pm x$ ($\pm y$)-valley hopping is longitudinal, and is therefore dominated by 
$\pm y$ ($\pm x$) and $\pm z$ valley components. 
It follows that the exchange energy is not sensitive to atomic scale position
variations along the direction of the array, since both $\hat{y} \cdot \bL $ ($\hat{x} \cdot \bL $) and $\hat{z} \cdot \bL $ are 0. However, there is a strong sensitivity to donor position variations in the direction perpendicular to donor array. 
In the $\langle 110 \rangle$ direction, the sensitivity of exchange interactions to donor placement 
is reduced in all in-plane directions since both $\pm x$ and $\pm y$ valley hopping strengths are reduced relative to
$\pm z$ valley hopping.  These results are 
consistent with recent experiments which have demonstrated a valley filtering effect on exchange interactions\cite{Salfi2018,Voisin2020}, which is captured with a very simple model in this paper.
As we explain in more detail in the next section, the lowest energy 
states become more concentrated in $A_1$ valley-split states 
at large donor separation, where $\epsilon_{vs}$ is larger than the hopping energies.
The problem of exchange splitting has been considered previously \cite{Koiller2001}
using the approximation, valid at large $L$, that the impurity-state Hamiltonian 
can be projected onto the $A_1$ basis, with qualitatively similar results for the large $L$ limit.

\textit{Disorder and Localization---}
Donor placement in silicon often errs by a lattice constant or more.
Even when a regular one-dimensional donor array is intended, the actual positions are $ R_i = (n_i N_i + \delta_i) a \hat{r_i}$, where $ \delta_i = 0, \, \pm 1$ randomly. 
Here $i = x,y,z$, $\hat{r_i}= \hat{x},\hat{y},\hat{z}$ are unit vectors along cubic axes, and $N_i$ is the intended superlattice length in units of the silicon lattice constant $a$. 
As we see from Eq.~\ref{eq:Hvs}, the random displacements introduce 
random phase factors $\exp{i(\bK_{\mu}-\bK_{\mu'}) \cdot \bR}$
in the off-diagonal matrix elements of the valley-splitting Hamiltonian. (See \cite{supplement} for further detail.)
These phase factors account for changes in the positions at which the system gains energy by 
establishing constructive interference between valleys.  
The phase factors are sensitive to atomic scale placement inaccuracy 
because valley momenta are comparable in size to microscopic silicon primitive reciprocal lattice vectors,
and much larger than the donor array superlattice primitive reciprocal lattice vectors. 
To study the influence of donor positional disorder on electronic properties 
we neglect interactions and calculate localization lengths using transfer matrices \cite{Beenakker1997} for a model in which the $\exp(i\bK_{\mu} \cdot \bR)$ ($\mu=x,y$) factors 
are modelled as independent random phase factors with phases $\Phi_{\mu}$.
$\Phi_z=0$ because the vertical component of the donor position is not expected to be disordered.

\ifpdf
\begin{figure}
  \centering
  \includegraphics[width=0.9 \linewidth ]{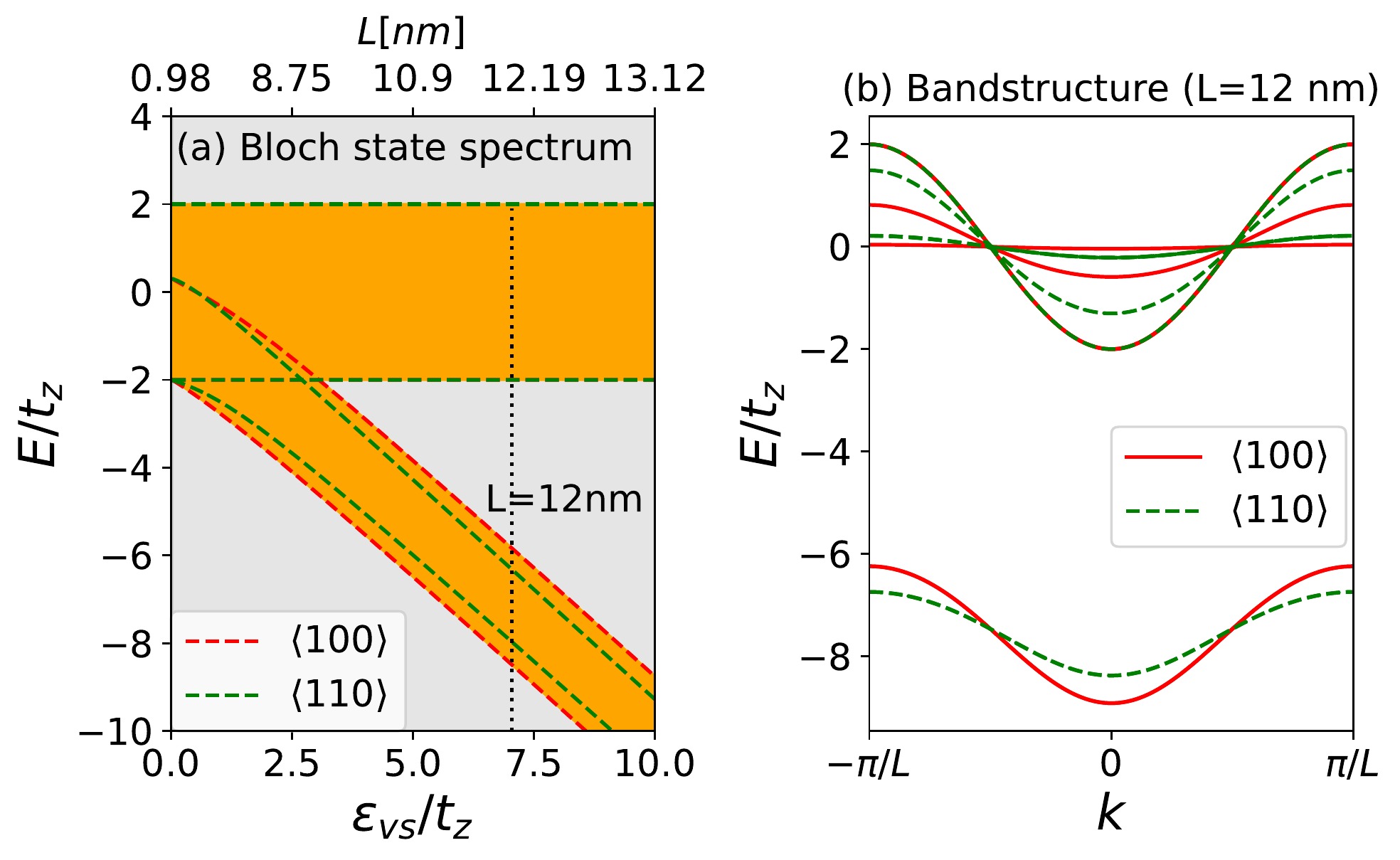}
  \caption{
  (a) Bloch state spectrum {\it vs.} valley splitting strength $\epsilon_{vs}=12$meV.  Energies are  
  in units of the hopping energy $t_z$.  The corresponding donor separation
  $L$ is indicated along the upper horizontal axis.  The green and red dashed lines specify the band edges 
  for donors placed along the $\langle 110 \rangle$ and $\langle 100 \rangle$ directions respectively and the orange regions are inside at least one of the six donor array bands for one of the two orientations. 
  (b) quasi-1D bandstructures for $L=12$nm donor arrays along $\langle 100 \rangle$ and 
  $\langle 110 \rangle$ directions respectively. The lowest energy (valley-split) $A_1$ band is singly degenerate in both cases.  In the $\langle 100 \rangle$ ($\langle 100 \rangle$) case
  the top dispersive (middle flatter) band with width of $4t_z$ ($4t_{x/y}$) has degeneracy of $3$. 
  }\label{Localization_no_disorder}
\end{figure}
\fi

The Bloch state spectrum of donor arrays placed in $\langle 100 \rangle$ and $\langle 110 \rangle$ directions are shown in Fig. \ref{Localization_no_disorder} (a). In the $L=12$ nm case, illustrated in Fig. \ref{Localization_no_disorder} (b), we see that the $A_1$ subband (lowered by $\epsilon_{vs}$) is splitted out, and that the width of this subband corresponds to an effective hopping amplitude that 
is intermediate between the longitudinal and $\hat{z}$-direction values, which will be discussed further below.

The donor array orientation dependence of our results is most easily understood in the large 
$\epsilon_{vs}$ limit, where we can truncate the Hilbert space to the $A_{1}$ donor levels.
The amplitude for hopping between $A_{1}$ levels is 
\begin{equation} 
t_{A1} = \frac{1}{6} \sum_{\mu} t_{\mu} \exp(i\Phi_{\mu})
\end{equation}
with $\Phi_z=0$ and $\Phi_{x}$ and $\Phi_y$ imposed by random $\hat{x}-\hat{y}$ plane positional shifts 
of the two donors.  Averaging over these phases we find that $\langle t_{A1} \rangle = t_z/3 $ and the 
coefficient of variation (the square root of the variance divided by the mean) is 
\begin{equation}
C_{A1}^v = \frac{\sigma_{A1}}{\langle t_{A1} \rangle} = \frac{\sqrt{t_x^2+t_y^2}}{t_z}.
\end{equation}
Here $\sigma_{A1}^2$ is the variance of $t_{A1}$ and $\langle t_{A1} \rangle$ is its average value.
The coefficient of variation for $t_{A1}$ is dominated by the larger of $t_x$ and $t_y$, and therefore 
reaches a minimum when $\theta = \pi/4$ since both $t_x$ and $t_y$ are reduced compared to 
$t_z$ in this case. In Fig.~\ref{Hopping_variance} (a) we plot $C_{A1}^v$ $\textit{vs.}$ $\theta$ for a series of $L$ values.  Here we see that the coefficient of variation of the hopping amplitude is close to $1$ at 
$\theta=0$, but reaches a minimum that drops with donor separation $L$ at $\theta=\pi/4$.
Hopping disorder is weaker for donor arrays aligned along $\langle 110 \rangle$ directions. 
         
As a result of this reduced disorder, the localization lengths of the 
$A_1$ band electrons reaches a maximum at $\theta = \pi/4$.
The localization lengths are 
defined as the inverse of the smallest positive Lyapunov exponent averaged over phase 
disorder realizations: $\xi_{loc} = \langle \gamma_m \rangle_{_{avg.}}^{-1}$ 
(see supplemental material \cite{supplement} for further detail).
In Fig. \ref{Hopping_variance} (b) we plot localization lengths calculated at an 
energy $0.5 t_z$ from the center of the disordered $A_1$ band, with 
$\Phi_{x/y}$ sampled from the interval $(-\Phi_m, \Phi_m)$ and $\Phi_m = \pi/2$, we see that the localization lengths maximize at $\langle 110 \rangle$ direction.  The $A_1$ band localization lengths are 
plotted {\it vs. } energy and $\Phi_m$ for $L=12$ nm in Fig. \ref{Hopping_variance} (c) and (d) for 
$\langle 100 \rangle$ and $\langle 110 \rangle$ directions respectively.  
for $\langle 110 \rangle$-orientation donor arrays, localization lengths remain long near the 
center of the $A_1$-band, even for strong donor-placement disorder.  

\ifpdf
\begin{figure}
  \centering
  \includegraphics[width=0.9 \linewidth ]{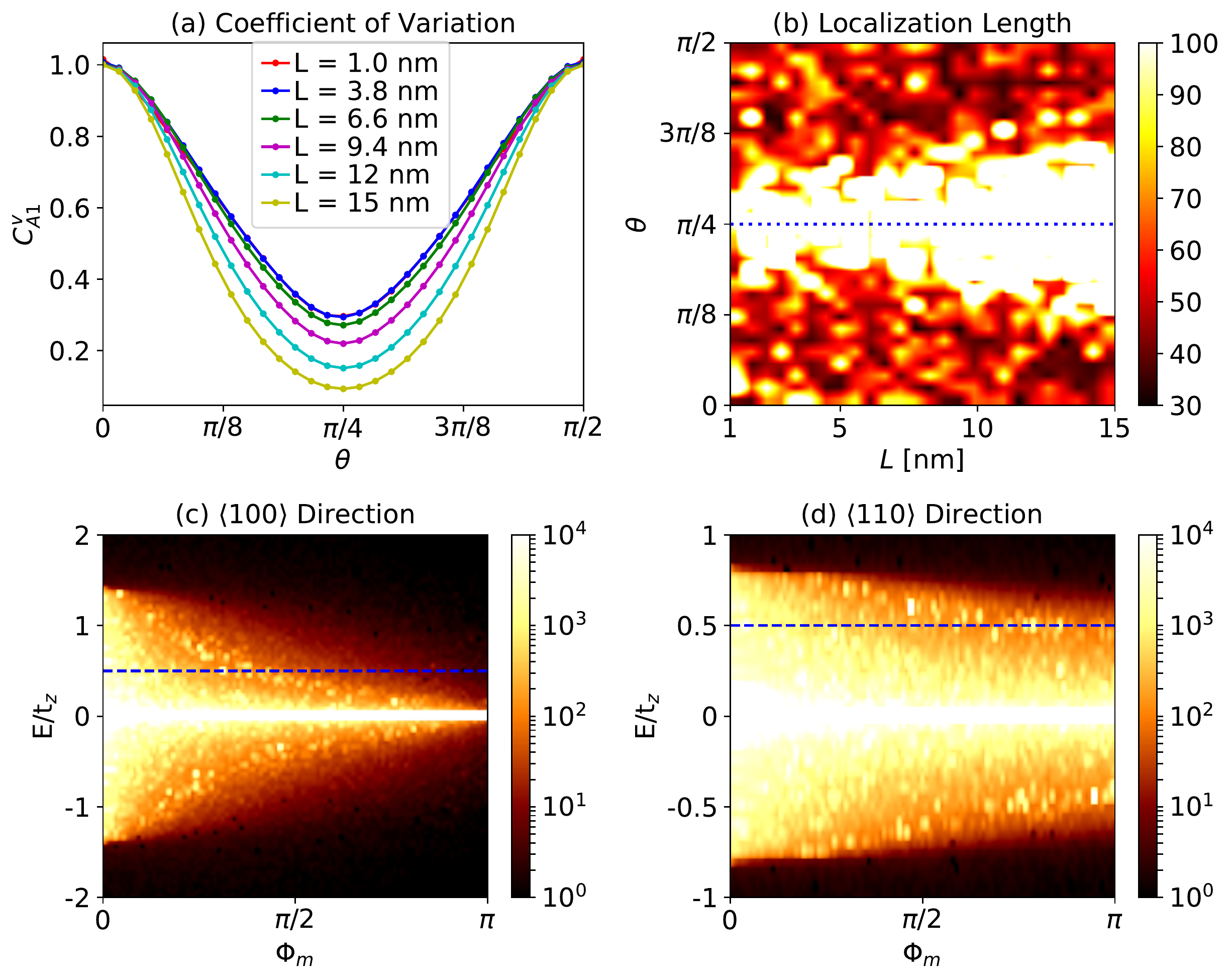}
  \caption{
  (a) Coefficient of variation for hopping between $A_1$ states. 
  Note that disorder is increasingly minimized at $\theta=\pi/4$ as $L$ increases.
  (b) Localization in donor separation units {\it vs.} donor separation $L$ and the 
  orientation of the donor array relative to the x-axis  $\theta$
  at an energy $E=0.5t_z$ above the middle of the disordered $A_1$ band. 
  (c) and (d) illustrate the dependence of $A_1$ band localization length on disorder strength and energy for 
  $L=12$nm.  The dashed lines in (c) and (d) mark $E=0.5t_z$, the energy used in (b).
  }\label{Hopping_variance}
\end{figure}
\fi

\textit{Discussions ---}
In summary, we have presented a theoretical model for a one-dimensional electron gas formed from donor arrays in silicon
that accounts in a simple but accurate way for valley-dependent central-cell corrections to the donor energy.
The model is based on the observations that there is an increase in donor binding energy $\epsilon_{vs}$ when all valley 
components of the donor wavefunction are in phase at the donor site.
For donor separations less than $\sim 4$nm, all six silicon valleys play an important role in low-energy
many-electron states.  At larger $L$, valley-splitting exceeds hopping energies and the low-energy physics 
can be approximated by a $A_1$ one-band model.  
Because the valley wavevectors are well separated in silicon, valley splitting leads to 
strong disorder in the hopping amplitude between the most strongly bound 
donor states, even when inaccuracy in donor placement is only at the microscopic silicon 
lattice constant scale.  It follows from our model that silicon donor arrays provide 
an excellent platform to study the combined influence of strong interactions and strong disorder in one-dimension.

The importance of interactions in silicon donor array states 
can be judged by evaluating the on-site Hubbard $U$.  
$U \sim 10$meV exceeds the hopping parameters at larger donor separations, reaching a 
ratio of $\sim 10$ when the donor separation is around $12$ nm.  In this limit only spin-degrees of 
freedom are relevant, donor arrays are thus a platform to study the physics in one or two dimensional disordered many-body system, such as random field Heisenberg model \cite{rfh_mbl} and one-dimensional Hubbard physics with hopping disorder \cite{Sandvik1994,Bloch1972,Giamarchi1995}.  In the limit of large $L$ disorder can be weakened, but not eliminated,
by orienting the donor array along the $\langle 110 \rangle$-direction. 

\textit{Acknowledgements---}  This work was supported by the Army Research Office (ARO)
under contract W911NF-15-1-0561:P00001.
We acknowledge the generous HPC resources provided by Texas Advanced Computing Center (TACC) 
at the University of Texas at Austin.

\bibliography{Donor}

\begin{thebibliography}{51}%
\makeatletter
\providecommand \@ifxundefined [1]{%
 \@ifx{#1\undefined}
}%
\providecommand \@ifnum [1]{%
 \ifnum #1\expandafter \@firstoftwo
 \else \expandafter \@secondoftwo
 \fi
}%
\providecommand \@ifx [1]{%
 \ifx #1\expandafter \@firstoftwo
 \else \expandafter \@secondoftwo
 \fi
}%
\providecommand \natexlab [1]{#1}%
\providecommand \enquote  [1]{``#1''}%
\providecommand \bibnamefont  [1]{#1}%
\providecommand \bibfnamefont [1]{#1}%
\providecommand \citenamefont [1]{#1}%
\providecommand \href@noop [0]{\@secondoftwo}%
\providecommand \href [0]{\begingroup \@sanitize@url \@href}%
\providecommand \@href[1]{\@@startlink{#1}\@@href}%
\providecommand \@@href[1]{\endgroup#1\@@endlink}%
\providecommand \@sanitize@url [0]{\catcode `\\12\catcode `\$12\catcode
  `\&12\catcode `\#12\catcode `\^12\catcode `\_12\catcode `\%12\relax}%
\providecommand \@@startlink[1]{}%
\providecommand \@@endlink[0]{}%
\providecommand \url  [0]{\begingroup\@sanitize@url \@url }%
\providecommand \@url [1]{\endgroup\@href {#1}{\urlprefix }}%
\providecommand \urlprefix  [0]{URL }%
\providecommand \Eprint [0]{\href }%
\providecommand \doibase [0]{http://dx.doi.org/}%
\providecommand \selectlanguage [0]{\@gobble}%
\providecommand \bibinfo  [0]{\@secondoftwo}%
\providecommand \bibfield  [0]{\@secondoftwo}%
\providecommand \translation [1]{[#1]}%
\providecommand \BibitemOpen [0]{}%
\providecommand \bibitemStop [0]{}%
\providecommand \bibitemNoStop [0]{.\EOS\space}%
\providecommand \EOS [0]{\spacefactor3000\relax}%
\providecommand \BibitemShut  [1]{\csname bibitem#1\endcsname}%
\let\auto@bib@innerbib\@empty
\bibitem [{\citenamefont {Zwanenburg}\ \emph {et~al.}(2013)\citenamefont
  {Zwanenburg}, \citenamefont {Dzurak}, \citenamefont {Morello}, \citenamefont
  {Simmons}, \citenamefont {Hollenberg}, \citenamefont {Klimeck}, \citenamefont
  {Rogge}, \citenamefont {Coppersmith},\ and\ \citenamefont
  {Eriksson}}]{Rev_silicon2013}%
  \BibitemOpen
  \bibfield  {author} {\bibinfo {author} {\bibfnamefont {F.~A.}\ \bibnamefont
  {Zwanenburg}}, \bibinfo {author} {\bibfnamefont {A.~S.}\ \bibnamefont
  {Dzurak}}, \bibinfo {author} {\bibfnamefont {A.}~\bibnamefont {Morello}},
  \bibinfo {author} {\bibfnamefont {M.~Y.}\ \bibnamefont {Simmons}}, \bibinfo
  {author} {\bibfnamefont {L.~C.~L.}\ \bibnamefont {Hollenberg}}, \bibinfo
  {author} {\bibfnamefont {G.}~\bibnamefont {Klimeck}}, \bibinfo {author}
  {\bibfnamefont {S.}~\bibnamefont {Rogge}}, \bibinfo {author} {\bibfnamefont
  {S.~N.}\ \bibnamefont {Coppersmith}}, \ and\ \bibinfo {author} {\bibfnamefont
  {M.~A.}\ \bibnamefont {Eriksson}},\ }\href {\doibase
  10.1103/RevModPhys.85.961} {\bibfield  {journal} {\bibinfo  {journal} {Rev.
  Mod. Phys.}\ }\textbf {\bibinfo {volume} {85}},\ \bibinfo {pages} {961}
  (\bibinfo {year} {2013})}\BibitemShut {NoStop}%
\bibitem [{\citenamefont {Knight}\ \emph {et~al.}(2003)\citenamefont {Knight},
  \citenamefont {Hinds}, \citenamefont {Plenio}, \citenamefont {Clark},
  \citenamefont {Brenner}, \citenamefont {Buehler}, \citenamefont {Chan},
  \citenamefont {Curson}, \citenamefont {Dzurak}, \citenamefont {Gauja},
  \citenamefont {Goan}, \citenamefont {Greentree}, \citenamefont {Hallam},
  \citenamefont {Hamilton}, \citenamefont {Hollenberg}, \citenamefont
  {Jamieson}, \citenamefont {McCallum}, \citenamefont {Milburn}, \citenamefont
  {O'Brien}, \citenamefont {Oberbeck}, \citenamefont {Pakes}, \citenamefont
  {Prawer}, \citenamefont {Reilly}, \citenamefont {Ruess}, \citenamefont
  {Schofield}, \citenamefont {Simmons}, \citenamefont {Stanley}, \citenamefont
  {Starrett}, \citenamefont {Wellard},\ and\ \citenamefont
  {Yang}}]{Rev_DonorQC2003}%
  \BibitemOpen
  \bibfield  {author} {\bibinfo {author} {\bibfnamefont {P.~L.}\ \bibnamefont
  {Knight}}, \bibinfo {author} {\bibfnamefont {E.~A.}\ \bibnamefont {Hinds}},
  \bibinfo {author} {\bibfnamefont {M.~B.}\ \bibnamefont {Plenio}}, \bibinfo
  {author} {\bibfnamefont {R.~G.}\ \bibnamefont {Clark}}, \bibinfo {author}
  {\bibfnamefont {R.}~\bibnamefont {Brenner}}, \bibinfo {author} {\bibfnamefont
  {T.~M.}\ \bibnamefont {Buehler}}, \bibinfo {author} {\bibfnamefont
  {V.}~\bibnamefont {Chan}}, \bibinfo {author} {\bibfnamefont {N.~J.}\
  \bibnamefont {Curson}}, \bibinfo {author} {\bibfnamefont {A.~S.}\
  \bibnamefont {Dzurak}}, \bibinfo {author} {\bibfnamefont {E.}~\bibnamefont
  {Gauja}}, \bibinfo {author} {\bibfnamefont {H.~S.}\ \bibnamefont {Goan}},
  \bibinfo {author} {\bibfnamefont {A.~D.}\ \bibnamefont {Greentree}}, \bibinfo
  {author} {\bibfnamefont {T.}~\bibnamefont {Hallam}}, \bibinfo {author}
  {\bibfnamefont {A.~R.}\ \bibnamefont {Hamilton}}, \bibinfo {author}
  {\bibfnamefont {L.~C.~L.}\ \bibnamefont {Hollenberg}}, \bibinfo {author}
  {\bibfnamefont {D.~N.}\ \bibnamefont {Jamieson}}, \bibinfo {author}
  {\bibfnamefont {J.~C.}\ \bibnamefont {McCallum}}, \bibinfo {author}
  {\bibfnamefont {G.~J.}\ \bibnamefont {Milburn}}, \bibinfo {author}
  {\bibfnamefont {J.~L.}\ \bibnamefont {O'Brien}}, \bibinfo {author}
  {\bibfnamefont {L.}~\bibnamefont {Oberbeck}}, \bibinfo {author}
  {\bibfnamefont {C.~I.}\ \bibnamefont {Pakes}}, \bibinfo {author}
  {\bibfnamefont {S.~D.}\ \bibnamefont {Prawer}}, \bibinfo {author}
  {\bibfnamefont {D.~J.}\ \bibnamefont {Reilly}}, \bibinfo {author}
  {\bibfnamefont {F.~J.}\ \bibnamefont {Ruess}}, \bibinfo {author}
  {\bibfnamefont {S.~R.}\ \bibnamefont {Schofield}}, \bibinfo {author}
  {\bibfnamefont {M.~Y.}\ \bibnamefont {Simmons}}, \bibinfo {author}
  {\bibfnamefont {F.~E.}\ \bibnamefont {Stanley}}, \bibinfo {author}
  {\bibfnamefont {R.~P.}\ \bibnamefont {Starrett}}, \bibinfo {author}
  {\bibfnamefont {C.}~\bibnamefont {Wellard}}, \ and\ \bibinfo {author}
  {\bibfnamefont {C.}~\bibnamefont {Yang}},\ }\href {\doibase
  10.1098/rsta.2003.1221} {\bibfield  {journal} {\bibinfo  {journal}
  {Philosophical Transactions of the Royal Society of London. Series A:
  Mathematical, Physical and Engineering Sciences}\ }\textbf {\bibinfo {volume}
  {361}},\ \bibinfo {pages} {1451} (\bibinfo {year} {2003})}\BibitemShut
  {NoStop}%
\bibitem [{\citenamefont {Pla}\ \emph {et~al.}(2012)\citenamefont {Pla},
  \citenamefont {Tan}, \citenamefont {Dehollain}, \citenamefont {Lim},
  \citenamefont {Morton}, \citenamefont {Jamieson}, \citenamefont {Dzurak},\
  and\ \citenamefont {Morello}}]{qubit2012}%
  \BibitemOpen
  \bibfield  {author} {\bibinfo {author} {\bibfnamefont {J.~J.}\ \bibnamefont
  {Pla}}, \bibinfo {author} {\bibfnamefont {K.~Y.}\ \bibnamefont {Tan}},
  \bibinfo {author} {\bibfnamefont {J.~P.}\ \bibnamefont {Dehollain}}, \bibinfo
  {author} {\bibfnamefont {W.~H.}\ \bibnamefont {Lim}}, \bibinfo {author}
  {\bibfnamefont {J.~J.~L.}\ \bibnamefont {Morton}}, \bibinfo {author}
  {\bibfnamefont {D.~N.}\ \bibnamefont {Jamieson}}, \bibinfo {author}
  {\bibfnamefont {A.~S.}\ \bibnamefont {Dzurak}}, \ and\ \bibinfo {author}
  {\bibfnamefont {A.}~\bibnamefont {Morello}},\ }\href {\doibase
  10.1038/nature11449} {\bibfield  {journal} {\bibinfo  {journal} {Nature}\
  }\textbf {\bibinfo {volume} {489}},\ \bibinfo {pages} {541} (\bibinfo {year}
  {2012})}\BibitemShut {NoStop}%
\bibitem [{\citenamefont {B\"{u}ch}\ \emph {et~al.}(2013)\citenamefont
  {B\"{u}ch}, \citenamefont {Mahapatra}, \citenamefont {Rahman}, \citenamefont
  {Morello},\ and\ \citenamefont {Simmons}}]{readout2013}%
  \BibitemOpen
  \bibfield  {author} {\bibinfo {author} {\bibfnamefont {H.}~\bibnamefont
  {B\"{u}ch}}, \bibinfo {author} {\bibfnamefont {S.}~\bibnamefont {Mahapatra}},
  \bibinfo {author} {\bibfnamefont {R.}~\bibnamefont {Rahman}}, \bibinfo
  {author} {\bibfnamefont {A.}~\bibnamefont {Morello}}, \ and\ \bibinfo
  {author} {\bibfnamefont {M.~Y.}\ \bibnamefont {Simmons}},\ }\href {\doibase
  10.1038/ncomms3017} {\bibfield  {journal} {\bibinfo  {journal} {Nat. Comm.}\
  }\textbf {\bibinfo {volume} {4}},\ \bibinfo {pages} {2017} (\bibinfo {year}
  {2013})}\BibitemShut {NoStop}%
\bibitem [{\citenamefont {Lo}\ \emph {et~al.}(2013)\citenamefont {Lo},
  \citenamefont {Weis}, \citenamefont {van Tol}, \citenamefont {Bokor},\ and\
  \citenamefont {Schenkel}}]{Lo2013}%
  \BibitemOpen
  \bibfield  {author} {\bibinfo {author} {\bibfnamefont {C.~C.}\ \bibnamefont
  {Lo}}, \bibinfo {author} {\bibfnamefont {C.~D.}\ \bibnamefont {Weis}},
  \bibinfo {author} {\bibfnamefont {J.}~\bibnamefont {van Tol}}, \bibinfo
  {author} {\bibfnamefont {J.}~\bibnamefont {Bokor}}, \ and\ \bibinfo {author}
  {\bibfnamefont {T.}~\bibnamefont {Schenkel}},\ }\href {\doibase
  10.1103/PhysRevLett.110.057601} {\bibfield  {journal} {\bibinfo  {journal}
  {Phys. Rev. Lett.}\ }\textbf {\bibinfo {volume} {110}},\ \bibinfo {pages}
  {057601} (\bibinfo {year} {2013})}\BibitemShut {NoStop}%
\bibitem [{\citenamefont {Pla}\ \emph {et~al.}(2014)\citenamefont {Pla},
  \citenamefont {Mohiyaddin}, \citenamefont {Tan}, \citenamefont {Dehollain},
  \citenamefont {Rahman}, \citenamefont {Klimeck}, \citenamefont {Jamieson},
  \citenamefont {Dzurak},\ and\ \citenamefont {Morello}}]{Pla2014}%
  \BibitemOpen
  \bibfield  {author} {\bibinfo {author} {\bibfnamefont {J.~J.}\ \bibnamefont
  {Pla}}, \bibinfo {author} {\bibfnamefont {F.~A.}\ \bibnamefont {Mohiyaddin}},
  \bibinfo {author} {\bibfnamefont {K.~Y.}\ \bibnamefont {Tan}}, \bibinfo
  {author} {\bibfnamefont {J.~P.}\ \bibnamefont {Dehollain}}, \bibinfo {author}
  {\bibfnamefont {R.}~\bibnamefont {Rahman}}, \bibinfo {author} {\bibfnamefont
  {G.}~\bibnamefont {Klimeck}}, \bibinfo {author} {\bibfnamefont {D.~N.}\
  \bibnamefont {Jamieson}}, \bibinfo {author} {\bibfnamefont {A.~S.}\
  \bibnamefont {Dzurak}}, \ and\ \bibinfo {author} {\bibfnamefont
  {A.}~\bibnamefont {Morello}},\ }\href {\doibase
  10.1103/PhysRevLett.113.246801} {\bibfield  {journal} {\bibinfo  {journal}
  {Phys. Rev. Lett.}\ }\textbf {\bibinfo {volume} {113}},\ \bibinfo {pages}
  {246801} (\bibinfo {year} {2014})}\BibitemShut {NoStop}%
\bibitem [{\citenamefont {Kalra}\ \emph {et~al.}(2014)\citenamefont {Kalra},
  \citenamefont {Laucht}, \citenamefont {Hill},\ and\ \citenamefont
  {Morello}}]{Kalra2014}%
  \BibitemOpen
  \bibfield  {author} {\bibinfo {author} {\bibfnamefont {R.}~\bibnamefont
  {Kalra}}, \bibinfo {author} {\bibfnamefont {A.}~\bibnamefont {Laucht}},
  \bibinfo {author} {\bibfnamefont {C.~D.}\ \bibnamefont {Hill}}, \ and\
  \bibinfo {author} {\bibfnamefont {A.}~\bibnamefont {Morello}},\ }\href
  {\doibase 10.1103/PhysRevX.4.021044} {\bibfield  {journal} {\bibinfo
  {journal} {Phys. Rev. X}\ }\textbf {\bibinfo {volume} {4}},\ \bibinfo {pages}
  {021044} (\bibinfo {year} {2014})}\BibitemShut {NoStop}%
\bibitem [{\citenamefont {Sigillito}\ \emph {et~al.}(2017)\citenamefont
  {Sigillito}, \citenamefont {Tyryshkin}, \citenamefont {Schenkel},
  \citenamefont {Houck},\ and\ \citenamefont {Lyon}}]{Sigillito2017}%
  \BibitemOpen
  \bibfield  {author} {\bibinfo {author} {\bibfnamefont {A.~J.}\ \bibnamefont
  {Sigillito}}, \bibinfo {author} {\bibfnamefont {A.~M.}\ \bibnamefont
  {Tyryshkin}}, \bibinfo {author} {\bibfnamefont {T.}~\bibnamefont {Schenkel}},
  \bibinfo {author} {\bibfnamefont {A.~A.}\ \bibnamefont {Houck}}, \ and\
  \bibinfo {author} {\bibfnamefont {S.~A.}\ \bibnamefont {Lyon}},\ }\href
  {\doibase 10.1038/nnano.2017.154} {\bibfield  {journal} {\bibinfo  {journal}
  {Nat. Nano.}\ }\textbf {\bibinfo {volume} {12}},\ \bibinfo {pages} {958}
  (\bibinfo {year} {2017})}\BibitemShut {NoStop}%
\bibitem [{\citenamefont {Veldhorst}\ \emph {et~al.}(2015)\citenamefont
  {Veldhorst}, \citenamefont {Yang}, \citenamefont {Hwang}, \citenamefont
  {Huang}, \citenamefont {Dehollain}, \citenamefont {Muhonen}, \citenamefont
  {Simmons}, \citenamefont {Laucht}, \citenamefont {Hudson}, \citenamefont
  {Itoh}, \citenamefont {Morello},\ and\ \citenamefont {Dzurak}}]{Gates2015}%
  \BibitemOpen
  \bibfield  {author} {\bibinfo {author} {\bibfnamefont {M.}~\bibnamefont
  {Veldhorst}}, \bibinfo {author} {\bibfnamefont {C.~H.}\ \bibnamefont {Yang}},
  \bibinfo {author} {\bibfnamefont {J.~C.~C.}\ \bibnamefont {Hwang}}, \bibinfo
  {author} {\bibfnamefont {W.}~\bibnamefont {Huang}}, \bibinfo {author}
  {\bibfnamefont {J.~P.}\ \bibnamefont {Dehollain}}, \bibinfo {author}
  {\bibfnamefont {J.~T.}\ \bibnamefont {Muhonen}}, \bibinfo {author}
  {\bibfnamefont {S.}~\bibnamefont {Simmons}}, \bibinfo {author} {\bibfnamefont
  {A.}~\bibnamefont {Laucht}}, \bibinfo {author} {\bibfnamefont {F.~E.}\
  \bibnamefont {Hudson}}, \bibinfo {author} {\bibfnamefont {K.~M.}\
  \bibnamefont {Itoh}}, \bibinfo {author} {\bibfnamefont {A.}~\bibnamefont
  {Morello}}, \ and\ \bibinfo {author} {\bibfnamefont {A.~S.}\ \bibnamefont
  {Dzurak}},\ }\href {\doibase 10.1038/nature15263} {\bibfield  {journal}
  {\bibinfo  {journal} {Nature}\ }\textbf {\bibinfo {volume} {526}},\ \bibinfo
  {pages} {410} (\bibinfo {year} {2015})}\BibitemShut {NoStop}%
\bibitem [{\citenamefont {He}\ \emph {et~al.}(2019)\citenamefont {He},
  \citenamefont {Gorman}, \citenamefont {Keith}, \citenamefont {Kranz},
  \citenamefont {Keizer},\ and\ \citenamefont {Simmons}}]{He2019}%
  \BibitemOpen
  \bibfield  {author} {\bibinfo {author} {\bibfnamefont {Y.}~\bibnamefont
  {He}}, \bibinfo {author} {\bibfnamefont {S.}~\bibnamefont {Gorman}}, \bibinfo
  {author} {\bibfnamefont {D.}~\bibnamefont {Keith}}, \bibinfo {author}
  {\bibfnamefont {L.}~\bibnamefont {Kranz}}, \bibinfo {author} {\bibfnamefont
  {J.}~\bibnamefont {Keizer}}, \ and\ \bibinfo {author} {\bibfnamefont
  {M.}~\bibnamefont {Simmons}},\ }\href {\doibase 10.1038/s41586-019-1381-2}
  {\bibfield  {journal} {\bibinfo  {journal} {Nature}\ }\textbf {\bibinfo
  {volume} {571}},\ \bibinfo {pages} {371} (\bibinfo {year}
  {2019})}\BibitemShut {NoStop}%
\bibitem [{\citenamefont {Morse}\ \emph {et~al.}(2017)\citenamefont {Morse},
  \citenamefont {Abraham}, \citenamefont {DeAbreu}, \citenamefont {Bowness},
  \citenamefont {Richards}, \citenamefont {Riemann}, \citenamefont {Abrosimov},
  \citenamefont {Becker}, \citenamefont {Pohl}, \citenamefont {Thewalt},\ and\
  \citenamefont {Simmons}}]{Morsee2017}%
  \BibitemOpen
  \bibfield  {author} {\bibinfo {author} {\bibfnamefont {K.~J.}\ \bibnamefont
  {Morse}}, \bibinfo {author} {\bibfnamefont {R.~J.~S.}\ \bibnamefont
  {Abraham}}, \bibinfo {author} {\bibfnamefont {A.}~\bibnamefont {DeAbreu}},
  \bibinfo {author} {\bibfnamefont {C.}~\bibnamefont {Bowness}}, \bibinfo
  {author} {\bibfnamefont {T.~S.}\ \bibnamefont {Richards}}, \bibinfo {author}
  {\bibfnamefont {H.}~\bibnamefont {Riemann}}, \bibinfo {author} {\bibfnamefont
  {N.~V.}\ \bibnamefont {Abrosimov}}, \bibinfo {author} {\bibfnamefont
  {P.}~\bibnamefont {Becker}}, \bibinfo {author} {\bibfnamefont {H.-J.}\
  \bibnamefont {Pohl}}, \bibinfo {author} {\bibfnamefont {M.~L.~W.}\
  \bibnamefont {Thewalt}}, \ and\ \bibinfo {author} {\bibfnamefont
  {S.}~\bibnamefont {Simmons}},\ }\href {\doibase 10.1126/sciadv.1700930}
  {\bibfield  {journal} {\bibinfo  {journal} {Science Advances}\ }\textbf
  {\bibinfo {volume} {3}} (\bibinfo {year} {2017}),\
  10.1126/sciadv.1700930}\BibitemShut {NoStop}%
\bibitem [{\citenamefont {Hile}\ \emph {et~al.}(2018)\citenamefont {Hile},
  \citenamefont {Fricke}, \citenamefont {House}, \citenamefont {Peretz},
  \citenamefont {Chen}, \citenamefont {Wang}, \citenamefont {Broome},
  \citenamefont {Gorman}, \citenamefont {Keizer}, \citenamefont {Rahman},\ and\
  \citenamefont {Simmons}}]{Hile2018}%
  \BibitemOpen
  \bibfield  {author} {\bibinfo {author} {\bibfnamefont {S.~J.}\ \bibnamefont
  {Hile}}, \bibinfo {author} {\bibfnamefont {L.}~\bibnamefont {Fricke}},
  \bibinfo {author} {\bibfnamefont {M.~G.}\ \bibnamefont {House}}, \bibinfo
  {author} {\bibfnamefont {E.}~\bibnamefont {Peretz}}, \bibinfo {author}
  {\bibfnamefont {C.~Y.}\ \bibnamefont {Chen}}, \bibinfo {author}
  {\bibfnamefont {Y.}~\bibnamefont {Wang}}, \bibinfo {author} {\bibfnamefont
  {M.}~\bibnamefont {Broome}}, \bibinfo {author} {\bibfnamefont {S.~K.}\
  \bibnamefont {Gorman}}, \bibinfo {author} {\bibfnamefont {J.~G.}\
  \bibnamefont {Keizer}}, \bibinfo {author} {\bibfnamefont {R.}~\bibnamefont
  {Rahman}}, \ and\ \bibinfo {author} {\bibfnamefont {M.~Y.}\ \bibnamefont
  {Simmons}},\ }\href {\doibase 10.1126/sciadv.aaq1459} {\bibfield  {journal}
  {\bibinfo  {journal} {Science Advances}\ }\textbf {\bibinfo {volume} {4}}
  (\bibinfo {year} {2018}),\ 10.1126/sciadv.aaq1459}\BibitemShut {NoStop}%
\bibitem [{\citenamefont {Dehollain}\ \emph {et~al.}(2014)\citenamefont
  {Dehollain}, \citenamefont {Muhonen}, \citenamefont {Tan}, \citenamefont
  {Saraiva}, \citenamefont {Jamieson}, \citenamefont {Dzurak},\ and\
  \citenamefont {Morello}}]{Dehollain2014}%
  \BibitemOpen
  \bibfield  {author} {\bibinfo {author} {\bibfnamefont {J.~P.}\ \bibnamefont
  {Dehollain}}, \bibinfo {author} {\bibfnamefont {J.~T.}\ \bibnamefont
  {Muhonen}}, \bibinfo {author} {\bibfnamefont {K.~Y.}\ \bibnamefont {Tan}},
  \bibinfo {author} {\bibfnamefont {A.}~\bibnamefont {Saraiva}}, \bibinfo
  {author} {\bibfnamefont {D.~N.}\ \bibnamefont {Jamieson}}, \bibinfo {author}
  {\bibfnamefont {A.~S.}\ \bibnamefont {Dzurak}}, \ and\ \bibinfo {author}
  {\bibfnamefont {A.}~\bibnamefont {Morello}},\ }\href {\doibase
  10.1103/PhysRevLett.112.236801} {\bibfield  {journal} {\bibinfo  {journal}
  {Phys. Rev. Lett.}\ }\textbf {\bibinfo {volume} {112}},\ \bibinfo {pages}
  {236801} (\bibinfo {year} {2014})}\BibitemShut {NoStop}%
\bibitem [{\citenamefont {Broome}\ \emph {et~al.}(2017)\citenamefont {Broome},
  \citenamefont {Watson}, \citenamefont {Keith}, \citenamefont {Gorman},
  \citenamefont {House}, \citenamefont {Keizer}, \citenamefont {Hile},
  \citenamefont {Baker},\ and\ \citenamefont {Simmons}}]{Broome2017}%
  \BibitemOpen
  \bibfield  {author} {\bibinfo {author} {\bibfnamefont {M.~A.}\ \bibnamefont
  {Broome}}, \bibinfo {author} {\bibfnamefont {T.~F.}\ \bibnamefont {Watson}},
  \bibinfo {author} {\bibfnamefont {D.}~\bibnamefont {Keith}}, \bibinfo
  {author} {\bibfnamefont {S.~K.}\ \bibnamefont {Gorman}}, \bibinfo {author}
  {\bibfnamefont {M.~G.}\ \bibnamefont {House}}, \bibinfo {author}
  {\bibfnamefont {J.~G.}\ \bibnamefont {Keizer}}, \bibinfo {author}
  {\bibfnamefont {S.~J.}\ \bibnamefont {Hile}}, \bibinfo {author}
  {\bibfnamefont {W.}~\bibnamefont {Baker}}, \ and\ \bibinfo {author}
  {\bibfnamefont {M.~Y.}\ \bibnamefont {Simmons}},\ }\href {\doibase
  10.1103/PhysRevLett.119.046802} {\bibfield  {journal} {\bibinfo  {journal}
  {Phys. Rev. Lett.}\ }\textbf {\bibinfo {volume} {119}},\ \bibinfo {pages}
  {046802} (\bibinfo {year} {2017})}\BibitemShut {NoStop}%
\bibitem [{\citenamefont {Zajac}\ \emph {et~al.}(2018)\citenamefont {Zajac},
  \citenamefont {Sigillito}, \citenamefont {Russ}, \citenamefont {Borjans},
  \citenamefont {Taylor}, \citenamefont {Burkard},\ and\ \citenamefont
  {Petta}}]{Zajac2018}%
  \BibitemOpen
  \bibfield  {author} {\bibinfo {author} {\bibfnamefont {D.~M.}\ \bibnamefont
  {Zajac}}, \bibinfo {author} {\bibfnamefont {A.~J.}\ \bibnamefont
  {Sigillito}}, \bibinfo {author} {\bibfnamefont {M.}~\bibnamefont {Russ}},
  \bibinfo {author} {\bibfnamefont {F.}~\bibnamefont {Borjans}}, \bibinfo
  {author} {\bibfnamefont {J.~M.}\ \bibnamefont {Taylor}}, \bibinfo {author}
  {\bibfnamefont {G.}~\bibnamefont {Burkard}}, \ and\ \bibinfo {author}
  {\bibfnamefont {J.~R.}\ \bibnamefont {Petta}},\ }\href {\doibase
  10.1126/science.aao5965} {\bibfield  {journal} {\bibinfo  {journal}
  {Science}\ }\textbf {\bibinfo {volume} {359}},\ \bibinfo {pages} {439}
  (\bibinfo {year} {2018})}\BibitemShut {NoStop}%
\bibitem [{\citenamefont {Watson}\ \emph {et~al.}(2018)\citenamefont {Watson},
  \citenamefont {Philips}, \citenamefont {Kawakami}, \citenamefont {Ward},
  \citenamefont {Scarlino}, \citenamefont {Veldhorst}, \citenamefont {Savage},
  \citenamefont {Lagally}, \citenamefont {Friesen}, \citenamefont
  {Coppersmith},\ and\ \citenamefont {et~al.}}]{Watson2018}%
  \BibitemOpen
  \bibfield  {author} {\bibinfo {author} {\bibfnamefont {T.~F.}\ \bibnamefont
  {Watson}}, \bibinfo {author} {\bibfnamefont {S.~G.~J.}\ \bibnamefont
  {Philips}}, \bibinfo {author} {\bibfnamefont {E.}~\bibnamefont {Kawakami}},
  \bibinfo {author} {\bibfnamefont {D.~R.}\ \bibnamefont {Ward}}, \bibinfo
  {author} {\bibfnamefont {P.}~\bibnamefont {Scarlino}}, \bibinfo {author}
  {\bibfnamefont {M.}~\bibnamefont {Veldhorst}}, \bibinfo {author}
  {\bibfnamefont {D.~E.}\ \bibnamefont {Savage}}, \bibinfo {author}
  {\bibfnamefont {M.~G.}\ \bibnamefont {Lagally}}, \bibinfo {author}
  {\bibfnamefont {M.}~\bibnamefont {Friesen}}, \bibinfo {author} {\bibfnamefont
  {S.~N.}\ \bibnamefont {Coppersmith}}, \ and\ \bibinfo {author} {\bibnamefont
  {et~al.}},\ }\href {\doibase 10.1038/nature25766} {\bibfield  {journal}
  {\bibinfo  {journal} {Nature}\ }\textbf {\bibinfo {volume} {555}},\ \bibinfo
  {pages} {633–637} (\bibinfo {year} {2018})}\BibitemShut {NoStop}%
\bibitem [{\citenamefont {Broome}\ \emph {et~al.}(2018)\citenamefont {Broome},
  \citenamefont {Gorman}, \citenamefont {House}, \citenamefont {Hile},
  \citenamefont {Keizer}, \citenamefont {Keith}, \citenamefont {Hill},
  \citenamefont {Watson}, \citenamefont {Baker}, \citenamefont {Hollenberg},\
  and\ \citenamefont {et~al.}}]{Broome2018}%
  \BibitemOpen
  \bibfield  {author} {\bibinfo {author} {\bibfnamefont {M.~A.}\ \bibnamefont
  {Broome}}, \bibinfo {author} {\bibfnamefont {S.~K.}\ \bibnamefont {Gorman}},
  \bibinfo {author} {\bibfnamefont {M.~G.}\ \bibnamefont {House}}, \bibinfo
  {author} {\bibfnamefont {S.~J.}\ \bibnamefont {Hile}}, \bibinfo {author}
  {\bibfnamefont {J.~G.}\ \bibnamefont {Keizer}}, \bibinfo {author}
  {\bibfnamefont {D.}~\bibnamefont {Keith}}, \bibinfo {author} {\bibfnamefont
  {C.~D.}\ \bibnamefont {Hill}}, \bibinfo {author} {\bibfnamefont {T.~F.}\
  \bibnamefont {Watson}}, \bibinfo {author} {\bibfnamefont {W.~J.}\
  \bibnamefont {Baker}}, \bibinfo {author} {\bibfnamefont {L.~C.~L.}\
  \bibnamefont {Hollenberg}}, \ and\ \bibinfo {author} {\bibnamefont
  {et~al.}},\ }\href {\doibase 10.1038/s41467-018-02982-x} {\bibfield
  {journal} {\bibinfo  {journal} {Nature Communications}\ }\textbf {\bibinfo
  {volume} {9}} (\bibinfo {year} {2018}),\
  10.1038/s41467-018-02982-x}\BibitemShut {NoStop}%
\bibitem [{\citenamefont {Wang}\ \emph {et~al.}(2016)\citenamefont {Wang},
  \citenamefont {Tankasala}, \citenamefont {Hollenberg}, \citenamefont
  {Klimeck}, \citenamefont {Simmons},\ and\ \citenamefont {Rahman}}]{Wang2016}%
  \BibitemOpen
  \bibfield  {author} {\bibinfo {author} {\bibfnamefont {Y.}~\bibnamefont
  {Wang}}, \bibinfo {author} {\bibfnamefont {A.}~\bibnamefont {Tankasala}},
  \bibinfo {author} {\bibfnamefont {L.~C.~L.}\ \bibnamefont {Hollenberg}},
  \bibinfo {author} {\bibfnamefont {G.}~\bibnamefont {Klimeck}}, \bibinfo
  {author} {\bibfnamefont {M.~Y.}\ \bibnamefont {Simmons}}, \ and\ \bibinfo
  {author} {\bibfnamefont {R.}~\bibnamefont {Rahman}},\ }\href {\doibase
  10.1038/npjqi.2016.8} {\bibfield  {journal} {\bibinfo  {journal} {npj Quantum
  Information}\ }\textbf {\bibinfo {volume} {2}} (\bibinfo {year} {2016}),\
  10.1038/npjqi.2016.8}\BibitemShut {NoStop}%
\bibitem [{\citenamefont {Tyryshkin}\ \emph {et~al.}(2011)\citenamefont
  {Tyryshkin}, \citenamefont {Tojo}, \citenamefont {Morton}, \citenamefont
  {Riemann}, \citenamefont {Abrosimov}, \citenamefont {Becker}, \citenamefont
  {Pohl}, \citenamefont {Schenkel}, \citenamefont {Thewalt}, \citenamefont
  {Itoh},\ and\ \citenamefont {Lyon}}]{coherence_e2011}%
  \BibitemOpen
  \bibfield  {author} {\bibinfo {author} {\bibfnamefont {A.~M.}\ \bibnamefont
  {Tyryshkin}}, \bibinfo {author} {\bibfnamefont {S.}~\bibnamefont {Tojo}},
  \bibinfo {author} {\bibfnamefont {J.~J.~L.}\ \bibnamefont {Morton}}, \bibinfo
  {author} {\bibfnamefont {H.}~\bibnamefont {Riemann}}, \bibinfo {author}
  {\bibfnamefont {N.~V.}\ \bibnamefont {Abrosimov}}, \bibinfo {author}
  {\bibfnamefont {P.}~\bibnamefont {Becker}}, \bibinfo {author} {\bibfnamefont
  {H.-J.}\ \bibnamefont {Pohl}}, \bibinfo {author} {\bibfnamefont
  {T.}~\bibnamefont {Schenkel}}, \bibinfo {author} {\bibfnamefont {M.~L.~W.}\
  \bibnamefont {Thewalt}}, \bibinfo {author} {\bibfnamefont {K.~M.}\
  \bibnamefont {Itoh}}, \ and\ \bibinfo {author} {\bibfnamefont {S.~A.}\
  \bibnamefont {Lyon}},\ }\href {\doibase 10.1038/nmat3182} {\bibfield
  {journal} {\bibinfo  {journal} {Nat. Mater.}\ }\textbf {\bibinfo {volume}
  {11}},\ \bibinfo {pages} {143} (\bibinfo {year} {2011})}\BibitemShut
  {NoStop}%
\bibitem [{\citenamefont {Watson}\ \emph {et~al.}(2017)\citenamefont {Watson},
  \citenamefont {Weber}, \citenamefont {Hsueh}, \citenamefont {Hollenberg},
  \citenamefont {Rahman},\ and\ \citenamefont {Simmons}}]{Watson2017}%
  \BibitemOpen
  \bibfield  {author} {\bibinfo {author} {\bibfnamefont {T.~F.}\ \bibnamefont
  {Watson}}, \bibinfo {author} {\bibfnamefont {B.}~\bibnamefont {Weber}},
  \bibinfo {author} {\bibfnamefont {Y.-L.}\ \bibnamefont {Hsueh}}, \bibinfo
  {author} {\bibfnamefont {L.~C.~L.}\ \bibnamefont {Hollenberg}}, \bibinfo
  {author} {\bibfnamefont {R.}~\bibnamefont {Rahman}}, \ and\ \bibinfo {author}
  {\bibfnamefont {M.~Y.}\ \bibnamefont {Simmons}},\ }\href {\doibase
  10.1126/sciadv.1602811} {\bibfield  {journal} {\bibinfo  {journal} {Science
  Advances}\ }\textbf {\bibinfo {volume} {3}} (\bibinfo {year} {2017}),\
  10.1126/sciadv.1602811}\BibitemShut {NoStop}%
\bibitem [{\citenamefont {Muhonen}\ \emph {et~al.}(2014)\citenamefont
  {Muhonen}, \citenamefont {Dehollain}, \citenamefont {Laucht}, \citenamefont
  {Hudson}, \citenamefont {Kalra}, \citenamefont {Sekiguchi}, \citenamefont
  {Itoh}, \citenamefont {Jamieson}, \citenamefont {McCallum}, \citenamefont
  {Dzurak},\ and\ \citenamefont {Morello}}]{coherence30s2014}%
  \BibitemOpen
  \bibfield  {author} {\bibinfo {author} {\bibfnamefont {J.~T.}\ \bibnamefont
  {Muhonen}}, \bibinfo {author} {\bibfnamefont {J.~P.}\ \bibnamefont
  {Dehollain}}, \bibinfo {author} {\bibfnamefont {A.}~\bibnamefont {Laucht}},
  \bibinfo {author} {\bibfnamefont {F.~E.}\ \bibnamefont {Hudson}}, \bibinfo
  {author} {\bibfnamefont {R.}~\bibnamefont {Kalra}}, \bibinfo {author}
  {\bibfnamefont {T.}~\bibnamefont {Sekiguchi}}, \bibinfo {author}
  {\bibfnamefont {K.~M.}\ \bibnamefont {Itoh}}, \bibinfo {author}
  {\bibfnamefont {D.~N.}\ \bibnamefont {Jamieson}}, \bibinfo {author}
  {\bibfnamefont {J.~C.}\ \bibnamefont {McCallum}}, \bibinfo {author}
  {\bibfnamefont {A.~S.}\ \bibnamefont {Dzurak}}, \ and\ \bibinfo {author}
  {\bibfnamefont {A.}~\bibnamefont {Morello}},\ }\href {\doibase
  10.1038/nnano.2014.211} {\bibfield  {journal} {\bibinfo  {journal} {Nat.
  Nano.}\ }\textbf {\bibinfo {volume} {9}},\ \bibinfo {pages} {986} (\bibinfo
  {year} {2014})}\BibitemShut {NoStop}%
\bibitem [{\citenamefont {Prati}\ \emph {et~al.}(2012)\citenamefont {Prati},
  \citenamefont {Hori}, \citenamefont {Guagliardo}, \citenamefont {Ferrari},\
  and\ \citenamefont {Shinada}}]{Mott2012}%
  \BibitemOpen
  \bibfield  {author} {\bibinfo {author} {\bibfnamefont {E.}~\bibnamefont
  {Prati}}, \bibinfo {author} {\bibfnamefont {M.}~\bibnamefont {Hori}},
  \bibinfo {author} {\bibfnamefont {F.}~\bibnamefont {Guagliardo}}, \bibinfo
  {author} {\bibfnamefont {G.}~\bibnamefont {Ferrari}}, \ and\ \bibinfo
  {author} {\bibfnamefont {T.}~\bibnamefont {Shinada}},\ }\href {\doibase
  10.1038/nnano.2012.94} {\bibfield  {journal} {\bibinfo  {journal} {Nat.
  Nano.}\ }\textbf {\bibinfo {volume} {7}},\ \bibinfo {pages} {443} (\bibinfo
  {year} {2012})}\BibitemShut {NoStop}%
\bibitem [{\citenamefont {Miwa}\ \emph {et~al.}(2013)\citenamefont {Miwa},
  \citenamefont {Hofmann}, \citenamefont {Simmons},\ and\ \citenamefont
  {Wells}}]{2D_gas2013}%
  \BibitemOpen
  \bibfield  {author} {\bibinfo {author} {\bibfnamefont {J.~A.}\ \bibnamefont
  {Miwa}}, \bibinfo {author} {\bibfnamefont {P.}~\bibnamefont {Hofmann}},
  \bibinfo {author} {\bibfnamefont {M.~Y.}\ \bibnamefont {Simmons}}, \ and\
  \bibinfo {author} {\bibfnamefont {J.~W.}\ \bibnamefont {Wells}},\ }\href
  {\doibase 10.1103/PhysRevLett.110.136801} {\bibfield  {journal} {\bibinfo
  {journal} {Phys. Rev. Lett.}\ }\textbf {\bibinfo {volume} {110}},\ \bibinfo
  {pages} {136801} (\bibinfo {year} {2013})}\BibitemShut {NoStop}%
\bibitem [{\citenamefont {Shamim}\ \emph {et~al.}(2014)\citenamefont {Shamim},
  \citenamefont {Mahapatra}, \citenamefont {Scappucci}, \citenamefont {Klesse},
  \citenamefont {Simmons},\ and\ \citenamefont {Ghosh}}]{donor2d2014}%
  \BibitemOpen
  \bibfield  {author} {\bibinfo {author} {\bibfnamefont {S.}~\bibnamefont
  {Shamim}}, \bibinfo {author} {\bibfnamefont {S.}~\bibnamefont {Mahapatra}},
  \bibinfo {author} {\bibfnamefont {G.}~\bibnamefont {Scappucci}}, \bibinfo
  {author} {\bibfnamefont {W.~M.}\ \bibnamefont {Klesse}}, \bibinfo {author}
  {\bibfnamefont {M.~Y.}\ \bibnamefont {Simmons}}, \ and\ \bibinfo {author}
  {\bibfnamefont {A.}~\bibnamefont {Ghosh}},\ }\href {\doibase
  10.1103/PhysRevLett.112.236602} {\bibfield  {journal} {\bibinfo  {journal}
  {Phys. Rev. Lett.}\ }\textbf {\bibinfo {volume} {112}},\ \bibinfo {pages}
  {236602} (\bibinfo {year} {2014})}\BibitemShut {NoStop}%
\bibitem [{\citenamefont {Prati}\ \emph {et~al.}(2016)\citenamefont {Prati},
  \citenamefont {Kumagai}, \citenamefont {Hori},\ and\ \citenamefont
  {Shinada}}]{donorchain2016}%
  \BibitemOpen
  \bibfield  {author} {\bibinfo {author} {\bibfnamefont {E.}~\bibnamefont
  {Prati}}, \bibinfo {author} {\bibfnamefont {K.}~\bibnamefont {Kumagai}},
  \bibinfo {author} {\bibfnamefont {M.}~\bibnamefont {Hori}}, \ and\ \bibinfo
  {author} {\bibfnamefont {T.}~\bibnamefont {Shinada}},\ }\href {\doibase
  10.1038/srep19704} {\bibfield  {journal} {\bibinfo  {journal} {Sci. Rep.}\
  }\textbf {\bibinfo {volume} {6}},\ \bibinfo {pages} {19704} (\bibinfo {year}
  {2016})}\BibitemShut {NoStop}%
\bibitem [{\citenamefont {Cooil}\ \emph {et~al.}(2017)\citenamefont {Cooil},
  \citenamefont {Mazzola}, \citenamefont {Klemm}, \citenamefont {Peschel},
  \citenamefont {Niu}, \citenamefont {Zakharov}, \citenamefont {Simmons},
  \citenamefont {Schmidt}, \citenamefont {Evans}, \citenamefont {Miwa},\ and\
  \citenamefont {Wells}}]{Cooil2017}%
  \BibitemOpen
  \bibfield  {author} {\bibinfo {author} {\bibfnamefont {S.~P.}\ \bibnamefont
  {Cooil}}, \bibinfo {author} {\bibfnamefont {F.}~\bibnamefont {Mazzola}},
  \bibinfo {author} {\bibfnamefont {H.~W.}\ \bibnamefont {Klemm}}, \bibinfo
  {author} {\bibfnamefont {G.}~\bibnamefont {Peschel}}, \bibinfo {author}
  {\bibfnamefont {Y.~R.}\ \bibnamefont {Niu}}, \bibinfo {author} {\bibfnamefont
  {A.~A.}\ \bibnamefont {Zakharov}}, \bibinfo {author} {\bibfnamefont {M.~Y.}\
  \bibnamefont {Simmons}}, \bibinfo {author} {\bibfnamefont {T.}~\bibnamefont
  {Schmidt}}, \bibinfo {author} {\bibfnamefont {D.~A.}\ \bibnamefont {Evans}},
  \bibinfo {author} {\bibfnamefont {J.~A.}\ \bibnamefont {Miwa}}, \ and\
  \bibinfo {author} {\bibfnamefont {J.~W.}\ \bibnamefont {Wells}},\ }\href
  {\doibase 10.1021/acsnano.6b07359} {\bibfield  {journal} {\bibinfo  {journal}
  {ACS Nano}\ }\textbf {\bibinfo {volume} {11}},\ \bibinfo {pages} {1683}
  (\bibinfo {year} {2017})},\ \bibinfo {note} {pMID: 28182399}\BibitemShut
  {NoStop}%
\bibitem [{\citenamefont {Kane}(1998)}]{Kane1998}%
  \BibitemOpen
  \bibfield  {author} {\bibinfo {author} {\bibfnamefont {B.~E.}\ \bibnamefont
  {Kane}},\ }\href {\doibase 10.1038/30156} {\bibfield  {journal} {\bibinfo
  {journal} {Nature}\ }\textbf {\bibinfo {volume} {393}},\ \bibinfo {pages}
  {133} (\bibinfo {year} {1998})}\BibitemShut {NoStop}%
\bibitem [{\citenamefont {Salfi}\ \emph {et~al.}(2016)\citenamefont {Salfi},
  \citenamefont {Mol}, \citenamefont {Rahman}, \citenamefont {Klimeck},
  \citenamefont {Simmons}, \citenamefont {Hollenberg},\ and\ \citenamefont
  {Rogge}}]{Salfi2016}%
  \BibitemOpen
  \bibfield  {author} {\bibinfo {author} {\bibfnamefont {J.}~\bibnamefont
  {Salfi}}, \bibinfo {author} {\bibfnamefont {J.~A.}\ \bibnamefont {Mol}},
  \bibinfo {author} {\bibfnamefont {R.}~\bibnamefont {Rahman}}, \bibinfo
  {author} {\bibfnamefont {G.}~\bibnamefont {Klimeck}}, \bibinfo {author}
  {\bibfnamefont {M.~Y.}\ \bibnamefont {Simmons}}, \bibinfo {author}
  {\bibfnamefont {L.~C.~L.}\ \bibnamefont {Hollenberg}}, \ and\ \bibinfo
  {author} {\bibfnamefont {S.}~\bibnamefont {Rogge}},\ }\href {\doibase
  10.1038/ncomms11342} {\bibfield  {journal} {\bibinfo  {journal} {Nat. Comm.}\
  }\textbf {\bibinfo {volume} {7}},\ \bibinfo {pages} {11342} (\bibinfo {year}
  {2016})}\BibitemShut {NoStop}%
\bibitem [{\citenamefont {Ansaloni}\ \emph {et~al.}(2020)\citenamefont
  {Ansaloni}, \citenamefont {Chatterjee}, \citenamefont {Bohuslavskyi},
  \citenamefont {Bertrand}, \citenamefont {Hutin}, \citenamefont {Vinet},\ and\
  \citenamefont {Kuemmeth}}]{Ansaloni_2020}%
  \BibitemOpen
  \bibfield  {author} {\bibinfo {author} {\bibfnamefont {F.}~\bibnamefont
  {Ansaloni}}, \bibinfo {author} {\bibfnamefont {A.}~\bibnamefont
  {Chatterjee}}, \bibinfo {author} {\bibfnamefont {H.}~\bibnamefont
  {Bohuslavskyi}}, \bibinfo {author} {\bibfnamefont {B.}~\bibnamefont
  {Bertrand}}, \bibinfo {author} {\bibfnamefont {L.}~\bibnamefont {Hutin}},
  \bibinfo {author} {\bibfnamefont {M.}~\bibnamefont {Vinet}}, \ and\ \bibinfo
  {author} {\bibfnamefont {F.}~\bibnamefont {Kuemmeth}},\ }\href {\doibase
  10.1038/s41467-020-20280-3} {\bibfield  {journal} {\bibinfo  {journal}
  {Nature Communications}\ }\textbf {\bibinfo {volume} {11}} (\bibinfo {year}
  {2020}),\ 10.1038/s41467-020-20280-3}\BibitemShut {NoStop}%
\bibitem [{\citenamefont {Pajot}\ and\ \citenamefont
  {Clerjaud}(2012)}]{Valley_in_silicon2012}%
  \BibitemOpen
  \bibfield  {author} {\bibinfo {author} {\bibfnamefont {B.}~\bibnamefont
  {Pajot}}\ and\ \bibinfo {author} {\bibfnamefont {B.}~\bibnamefont
  {Clerjaud}},\ }\href@noop {} {\emph {\bibinfo {title} {Optical Absorption of
  Impurities and Defects in Semiconducting Crystals: Electronic Absorption of
  Deep Centres and Vibrational Spectra}}},\ Vol.\ \bibinfo {volume} {169}\
  (\bibinfo  {publisher} {Springer Science \& Business Media},\ \bibinfo {year}
  {2012})\BibitemShut {NoStop}%
\bibitem [{\citenamefont {Goh}\ \emph {et~al.}(2020)\citenamefont {Goh},
  \citenamefont {Bussolotti}, \citenamefont {Lau}, \citenamefont
  {Kotekar-Patil}, \citenamefont {Ooi},\ and\ \citenamefont {Chee}}]{Goh2020}%
  \BibitemOpen
  \bibfield  {author} {\bibinfo {author} {\bibfnamefont {K.~E.~J.}\
  \bibnamefont {Goh}}, \bibinfo {author} {\bibfnamefont {F.}~\bibnamefont
  {Bussolotti}}, \bibinfo {author} {\bibfnamefont {C.~S.}\ \bibnamefont {Lau}},
  \bibinfo {author} {\bibfnamefont {D.}~\bibnamefont {Kotekar-Patil}}, \bibinfo
  {author} {\bibfnamefont {Z.~E.}\ \bibnamefont {Ooi}}, \ and\ \bibinfo
  {author} {\bibfnamefont {J.~Y.}\ \bibnamefont {Chee}},\ }\href@noop {}
  {\bibfield  {journal} {\bibinfo  {journal} {Advanced Quantum Technologies}\
  }\textbf {\bibinfo {volume} {3}},\ \bibinfo {pages} {1900123} (\bibinfo
  {year} {2020})}\BibitemShut {NoStop}%
\bibitem [{\citenamefont {Schofield}\ \emph {et~al.}(2003)\citenamefont
  {Schofield}, \citenamefont {Curson}, \citenamefont {Simmons}, \citenamefont
  {Rue\ss{}}, \citenamefont {Hallam}, \citenamefont {Oberbeck},\ and\
  \citenamefont {Clark}}]{Precise2003}%
  \BibitemOpen
  \bibfield  {author} {\bibinfo {author} {\bibfnamefont {S.~R.}\ \bibnamefont
  {Schofield}}, \bibinfo {author} {\bibfnamefont {N.~J.}\ \bibnamefont
  {Curson}}, \bibinfo {author} {\bibfnamefont {M.~Y.}\ \bibnamefont {Simmons}},
  \bibinfo {author} {\bibfnamefont {F.~J.}\ \bibnamefont {Rue\ss{}}}, \bibinfo
  {author} {\bibfnamefont {T.}~\bibnamefont {Hallam}}, \bibinfo {author}
  {\bibfnamefont {L.}~\bibnamefont {Oberbeck}}, \ and\ \bibinfo {author}
  {\bibfnamefont {R.~G.}\ \bibnamefont {Clark}},\ }\href {\doibase
  10.1103/PhysRevLett.91.136104} {\bibfield  {journal} {\bibinfo  {journal}
  {Phys. Rev. Lett.}\ }\textbf {\bibinfo {volume} {91}},\ \bibinfo {pages}
  {136104} (\bibinfo {year} {2003})}\BibitemShut {NoStop}%
\bibitem [{\citenamefont {Salfi}\ \emph {et~al.}(2018)\citenamefont {Salfi},
  \citenamefont {Voisin}, \citenamefont {Tankasala}, \citenamefont {Bocquel},
  \citenamefont {Usman}, \citenamefont {Simmons}, \citenamefont {Hollenberg},
  \citenamefont {Rahman},\ and\ \citenamefont {Rogge}}]{Salfi2018}%
  \BibitemOpen
  \bibfield  {author} {\bibinfo {author} {\bibfnamefont {J.}~\bibnamefont
  {Salfi}}, \bibinfo {author} {\bibfnamefont {B.}~\bibnamefont {Voisin}},
  \bibinfo {author} {\bibfnamefont {A.}~\bibnamefont {Tankasala}}, \bibinfo
  {author} {\bibfnamefont {J.}~\bibnamefont {Bocquel}}, \bibinfo {author}
  {\bibfnamefont {M.}~\bibnamefont {Usman}}, \bibinfo {author} {\bibfnamefont
  {M.~Y.}\ \bibnamefont {Simmons}}, \bibinfo {author} {\bibfnamefont
  {L.~C.~L.}\ \bibnamefont {Hollenberg}}, \bibinfo {author} {\bibfnamefont
  {R.}~\bibnamefont {Rahman}}, \ and\ \bibinfo {author} {\bibfnamefont
  {S.}~\bibnamefont {Rogge}},\ }\href {\doibase 10.1103/PhysRevX.8.031049}
  {\bibfield  {journal} {\bibinfo  {journal} {Phys. Rev. X}\ }\textbf {\bibinfo
  {volume} {8}},\ \bibinfo {pages} {031049} (\bibinfo {year}
  {2018})}\BibitemShut {NoStop}%
\bibitem [{\citenamefont {Voisin}\ \emph {et~al.}(2020)\citenamefont {Voisin},
  \citenamefont {Bocquel}, \citenamefont {Tankasala}, \citenamefont {Usman},
  \citenamefont {Salfi}, \citenamefont {Rahman}, \citenamefont {Simmons},
  \citenamefont {Hollenberg},\ and\ \citenamefont {Rogge}}]{Voisin2020}%
  \BibitemOpen
  \bibfield  {author} {\bibinfo {author} {\bibfnamefont {B.}~\bibnamefont
  {Voisin}}, \bibinfo {author} {\bibfnamefont {J.}~\bibnamefont {Bocquel}},
  \bibinfo {author} {\bibfnamefont {A.}~\bibnamefont {Tankasala}}, \bibinfo
  {author} {\bibfnamefont {M.}~\bibnamefont {Usman}}, \bibinfo {author}
  {\bibfnamefont {J.}~\bibnamefont {Salfi}}, \bibinfo {author} {\bibfnamefont
  {R.}~\bibnamefont {Rahman}}, \bibinfo {author} {\bibfnamefont
  {M.}~\bibnamefont {Simmons}}, \bibinfo {author} {\bibfnamefont
  {L.}~\bibnamefont {Hollenberg}}, \ and\ \bibinfo {author} {\bibfnamefont
  {S.}~\bibnamefont {Rogge}},\ }\href@noop {} {\bibfield  {journal} {\bibinfo
  {journal} {Nature communications}\ }\textbf {\bibinfo {volume} {11}},\
  \bibinfo {pages} {1} (\bibinfo {year} {2020})}\BibitemShut {NoStop}%
\bibitem [{\citenamefont {Kohn}\ and\ \citenamefont
  {Luttinger}(1955)}]{Theory_Kohn1955}%
  \BibitemOpen
  \bibfield  {author} {\bibinfo {author} {\bibfnamefont {W.}~\bibnamefont
  {Kohn}}\ and\ \bibinfo {author} {\bibfnamefont {J.~M.}\ \bibnamefont
  {Luttinger}},\ }\href {\doibase 10.1103/PhysRev.98.915} {\bibfield  {journal}
  {\bibinfo  {journal} {Phys. Rev.}\ }\textbf {\bibinfo {volume} {98}},\
  \bibinfo {pages} {915} (\bibinfo {year} {1955})}\BibitemShut {NoStop}%
\bibitem [{\citenamefont {Luttinger}\ and\ \citenamefont
  {Kohn}(1955)}]{Theory_Luttinger1955}%
  \BibitemOpen
  \bibfield  {author} {\bibinfo {author} {\bibfnamefont {J.~M.}\ \bibnamefont
  {Luttinger}}\ and\ \bibinfo {author} {\bibfnamefont {W.}~\bibnamefont
  {Kohn}},\ }\href {\doibase 10.1103/PhysRev.97.869} {\bibfield  {journal}
  {\bibinfo  {journal} {Phys. Rev.}\ }\textbf {\bibinfo {volume} {97}},\
  \bibinfo {pages} {869} (\bibinfo {year} {1955})}\BibitemShut {NoStop}%
\bibitem [{\citenamefont {Kittel}\ and\ \citenamefont
  {Mitchell}(1954)}]{Theory_Kittel1954}%
  \BibitemOpen
  \bibfield  {author} {\bibinfo {author} {\bibfnamefont {C.}~\bibnamefont
  {Kittel}}\ and\ \bibinfo {author} {\bibfnamefont {A.~H.}\ \bibnamefont
  {Mitchell}},\ }\href {\doibase 10.1103/PhysRev.96.1488} {\bibfield  {journal}
  {\bibinfo  {journal} {Phys. Rev.}\ }\textbf {\bibinfo {volume} {96}},\
  \bibinfo {pages} {1488} (\bibinfo {year} {1954})}\BibitemShut {NoStop}%
\bibitem [{\citenamefont {Baldereschi}\ and\ \citenamefont
  {Lipari}(1973)}]{Shallow1973}%
  \BibitemOpen
  \bibfield  {author} {\bibinfo {author} {\bibfnamefont {A.}~\bibnamefont
  {Baldereschi}}\ and\ \bibinfo {author} {\bibfnamefont {N.~O.}\ \bibnamefont
  {Lipari}},\ }\href {\doibase 10.1103/PhysRevB.8.2697} {\bibfield  {journal}
  {\bibinfo  {journal} {Phys. Rev. B}\ }\textbf {\bibinfo {volume} {8}},\
  \bibinfo {pages} {2697} (\bibinfo {year} {1973})}\BibitemShut {NoStop}%
\bibitem [{\citenamefont {Shindo}\ and\ \citenamefont
  {Nara}(1976)}]{Shindo1976}%
  \BibitemOpen
  \bibfield  {author} {\bibinfo {author} {\bibfnamefont {K.}~\bibnamefont
  {Shindo}}\ and\ \bibinfo {author} {\bibfnamefont {H.}~\bibnamefont {Nara}},\
  }\href {\doibase 10.1143/JPSJ.40.1640} {\bibfield  {journal} {\bibinfo
  {journal} {J. Phys. Soc. Jpn.}\ }\textbf {\bibinfo {volume} {40}},\ \bibinfo
  {pages} {1640} (\bibinfo {year} {1976})}\BibitemShut {NoStop}%
\bibitem [{\citenamefont {Saraiva}\ \emph {et~al.}(2011)\citenamefont
  {Saraiva}, \citenamefont {Calder\'on}, \citenamefont {Capaz}, \citenamefont
  {Hu}, \citenamefont {Das~Sarma},\ and\ \citenamefont
  {Koiller}}]{Saraiva2011}%
  \BibitemOpen
  \bibfield  {author} {\bibinfo {author} {\bibfnamefont {A.~L.}\ \bibnamefont
  {Saraiva}}, \bibinfo {author} {\bibfnamefont {M.~J.}\ \bibnamefont
  {Calder\'on}}, \bibinfo {author} {\bibfnamefont {R.~B.}\ \bibnamefont
  {Capaz}}, \bibinfo {author} {\bibfnamefont {X.}~\bibnamefont {Hu}}, \bibinfo
  {author} {\bibfnamefont {S.}~\bibnamefont {Das~Sarma}}, \ and\ \bibinfo
  {author} {\bibfnamefont {B.}~\bibnamefont {Koiller}},\ }\href {\doibase
  10.1103/PhysRevB.84.155320} {\bibfield  {journal} {\bibinfo  {journal} {Phys.
  Rev. B}\ }\textbf {\bibinfo {volume} {84}},\ \bibinfo {pages} {155320}
  (\bibinfo {year} {2011})}\BibitemShut {NoStop}%
\bibitem [{\citenamefont {Giuliani}\ and\ \citenamefont
  {Vignale}(2005)}]{Giuliani2005}%
  \BibitemOpen
  \bibfield  {author} {\bibinfo {author} {\bibfnamefont {G.}~\bibnamefont
  {Giuliani}}\ and\ \bibinfo {author} {\bibfnamefont {G.}~\bibnamefont
  {Vignale}},\ }\href {\doibase 10.1017/CBO9780511619915} {\emph {\bibinfo
  {title} {Quantum theory of the electron liquid}}}\ (\bibinfo  {publisher}
  {Cambridge University Press},\ \bibinfo {year} {2005})\BibitemShut {NoStop}%
\bibitem [{\citenamefont {Ramdas}\ and\ \citenamefont
  {Rodriguez}(1981)}]{bindenergy1981}%
  \BibitemOpen
  \bibfield  {author} {\bibinfo {author} {\bibfnamefont {A.~K.}\ \bibnamefont
  {Ramdas}}\ and\ \bibinfo {author} {\bibfnamefont {S.}~\bibnamefont
  {Rodriguez}},\ }\href {\doibase 10.1088/0034-4885/44/12/002} {\bibfield
  {journal} {\bibinfo  {journal} {Reports on Progress in Physics}\ }\textbf
  {\bibinfo {volume} {44}},\ \bibinfo {pages} {1297} (\bibinfo {year}
  {1981})}\BibitemShut {NoStop}%
\bibitem [{\citenamefont {Mayur}\ \emph {et~al.}(1993)\citenamefont {Mayur},
  \citenamefont {Sciacca}, \citenamefont {Ramdas},\ and\ \citenamefont
  {Rodriguez}}]{bindenergy1993}%
  \BibitemOpen
  \bibfield  {author} {\bibinfo {author} {\bibfnamefont {A.~J.}\ \bibnamefont
  {Mayur}}, \bibinfo {author} {\bibfnamefont {M.~D.}\ \bibnamefont {Sciacca}},
  \bibinfo {author} {\bibfnamefont {A.~K.}\ \bibnamefont {Ramdas}}, \ and\
  \bibinfo {author} {\bibfnamefont {S.}~\bibnamefont {Rodriguez}},\ }\href
  {\doibase 10.1103/PhysRevB.48.10893} {\bibfield  {journal} {\bibinfo
  {journal} {Phys. Rev. B}\ }\textbf {\bibinfo {volume} {48}},\ \bibinfo
  {pages} {10893} (\bibinfo {year} {1993})}\BibitemShut {NoStop}%
\bibitem [{\citenamefont {Gamble}\ \emph {et~al.}(2015)\citenamefont {Gamble},
  \citenamefont {Jacobson}, \citenamefont {Nielsen}, \citenamefont {Baczewski},
  \citenamefont {Moussa}, \citenamefont {Monta\~no},\ and\ \citenamefont
  {Muller}}]{Multivalley2015}%
  \BibitemOpen
  \bibfield  {author} {\bibinfo {author} {\bibfnamefont {J.~K.}\ \bibnamefont
  {Gamble}}, \bibinfo {author} {\bibfnamefont {N.~T.}\ \bibnamefont
  {Jacobson}}, \bibinfo {author} {\bibfnamefont {E.}~\bibnamefont {Nielsen}},
  \bibinfo {author} {\bibfnamefont {A.~D.}\ \bibnamefont {Baczewski}}, \bibinfo
  {author} {\bibfnamefont {J.~E.}\ \bibnamefont {Moussa}}, \bibinfo {author}
  {\bibfnamefont {I.}~\bibnamefont {Monta\~no}}, \ and\ \bibinfo {author}
  {\bibfnamefont {R.~P.}\ \bibnamefont {Muller}},\ }\href {\doibase
  10.1103/PhysRevB.91.235318} {\bibfield  {journal} {\bibinfo  {journal} {Phys.
  Rev. B}\ }\textbf {\bibinfo {volume} {91}},\ \bibinfo {pages} {235318}
  (\bibinfo {year} {2015})}\BibitemShut {NoStop}%
\bibitem [{\citenamefont {Koiller}\ \emph {et~al.}(2001)\citenamefont
  {Koiller}, \citenamefont {Hu},\ and\ \citenamefont
  {Das~Sarma}}]{Koiller2001}%
  \BibitemOpen
  \bibfield  {author} {\bibinfo {author} {\bibfnamefont {B.}~\bibnamefont
  {Koiller}}, \bibinfo {author} {\bibfnamefont {X.}~\bibnamefont {Hu}}, \ and\
  \bibinfo {author} {\bibfnamefont {S.}~\bibnamefont {Das~Sarma}},\ }\href
  {\doibase 10.1103/PhysRevLett.88.027903} {\bibfield  {journal} {\bibinfo
  {journal} {Phys. Rev. Lett.}\ }\textbf {\bibinfo {volume} {88}},\ \bibinfo
  {pages} {027903} (\bibinfo {year} {2001})}\BibitemShut {NoStop}%
\bibitem [{sup()}]{supplement}%
  \BibitemOpen
  \href@noop {} {\bibinfo  {journal} {See Supplemental Material at [URL] for
  more information}\ }\BibitemShut {NoStop}%
\bibitem [{\citenamefont {Beenakker}(1997)}]{Beenakker1997}%
  \BibitemOpen
\bibfield  {journal} {  }\bibfield  {author} {\bibinfo {author} {\bibfnamefont
  {C.~W.~J.}\ \bibnamefont {Beenakker}},\ }\href {\doibase
  10.1103/RevModPhys.69.731} {\bibfield  {journal} {\bibinfo  {journal} {Rev.
  Mod. Phys.}\ }\textbf {\bibinfo {volume} {69}},\ \bibinfo {pages} {731}
  (\bibinfo {year} {1997})}\BibitemShut {NoStop}%
\bibitem [{\citenamefont {\ifmmode \check{Z}\else
  \v{Z}\fi{}nidari\ifmmode~\check{c}\else \v{c}\fi{}}\ \emph
  {et~al.}(2008)\citenamefont {\ifmmode \check{Z}\else
  \v{Z}\fi{}nidari\ifmmode~\check{c}\else \v{c}\fi{}}, \citenamefont {Prosen},\
  and\ \citenamefont {Prelov\ifmmode~\check{s}\else \v{s}\fi{}ek}}]{rfh_mbl}%
  \BibitemOpen
  \bibfield  {author} {\bibinfo {author} {\bibfnamefont {M.}~\bibnamefont
  {\ifmmode \check{Z}\else \v{Z}\fi{}nidari\ifmmode~\check{c}\else
  \v{c}\fi{}}}, \bibinfo {author} {\bibfnamefont {T.~c.~v.}\ \bibnamefont
  {Prosen}}, \ and\ \bibinfo {author} {\bibfnamefont {P.}~\bibnamefont
  {Prelov\ifmmode~\check{s}\else \v{s}\fi{}ek}},\ }\href {\doibase
  10.1103/PhysRevB.77.064426} {\bibfield  {journal} {\bibinfo  {journal} {Phys.
  Rev. B}\ }\textbf {\bibinfo {volume} {77}},\ \bibinfo {pages} {064426}
  (\bibinfo {year} {2008})}\BibitemShut {NoStop}%
\bibitem [{\citenamefont {Sandvik}\ \emph {et~al.}(1994)\citenamefont
  {Sandvik}, \citenamefont {Scalapino},\ and\ \citenamefont
  {Henelius}}]{Sandvik1994}%
  \BibitemOpen
  \bibfield  {author} {\bibinfo {author} {\bibfnamefont {A.~W.}\ \bibnamefont
  {Sandvik}}, \bibinfo {author} {\bibfnamefont {D.~J.}\ \bibnamefont
  {Scalapino}}, \ and\ \bibinfo {author} {\bibfnamefont {P.}~\bibnamefont
  {Henelius}},\ }\href {\doibase 10.1103/PhysRevB.50.10474} {\bibfield
  {journal} {\bibinfo  {journal} {Phys. Rev. B}\ }\textbf {\bibinfo {volume}
  {50}},\ \bibinfo {pages} {10474} (\bibinfo {year} {1994})}\BibitemShut
  {NoStop}%
\bibitem [{\citenamefont {Bloch}\ \emph {et~al.}(1972)\citenamefont {Bloch},
  \citenamefont {Weisman},\ and\ \citenamefont {Varma}}]{Bloch1972}%
  \BibitemOpen
  \bibfield  {author} {\bibinfo {author} {\bibfnamefont {A.~N.}\ \bibnamefont
  {Bloch}}, \bibinfo {author} {\bibfnamefont {R.~B.}\ \bibnamefont {Weisman}},
  \ and\ \bibinfo {author} {\bibfnamefont {C.~M.}\ \bibnamefont {Varma}},\
  }\href {\doibase 10.1103/PhysRevLett.28.753} {\bibfield  {journal} {\bibinfo
  {journal} {Phys. Rev. Lett.}\ }\textbf {\bibinfo {volume} {28}},\ \bibinfo
  {pages} {753} (\bibinfo {year} {1972})}\BibitemShut {NoStop}%
\bibitem [{\citenamefont {Giamarchi}\ and\ \citenamefont
  {Shastry}(1995)}]{Giamarchi1995}%
  \BibitemOpen
  \bibfield  {author} {\bibinfo {author} {\bibfnamefont {T.}~\bibnamefont
  {Giamarchi}}\ and\ \bibinfo {author} {\bibfnamefont {B.~S.}\ \bibnamefont
  {Shastry}},\ }\href {\doibase 10.1103/PhysRevB.51.10915} {\bibfield
  {journal} {\bibinfo  {journal} {Phys. Rev. B}\ }\textbf {\bibinfo {volume}
  {51}},\ \bibinfo {pages} {10915} (\bibinfo {year} {1995})}\BibitemShut
  {NoStop}%
\end{thebibliography}%

\end{document}